\documentclass[aps,prr,amsmath,amssymb,floatfix,twocolumn]{revtex4-1}

\usepackage{multirow}
\usepackage{bbold}
\usepackage{graphicx}
\usepackage{subfigure}
\usepackage{color}
\usepackage{hyperref}
\usepackage[justification=raggedright]{caption}
\usepackage{filecontents}

\begin{document}
\preprint{APS/123-QED}

\title{Fine tuning generative adversarial networks with universal force fields: application to two-dimensional topological insulators}
\author{Alexander C. Tyner$^{1,2}$}

\affiliation{$^{1}$ NORDITA, KTH Royal Institute of Technology and Stockholm University 106 91 Stockholm, Sweden}
\affiliation{$^2$ Department of Physics, University of Connecticut, Storrs, Connecticut 06269, USA}

\date{\today}

\begin{abstract} 
Despite rapid growth in use cases for generative artificial intelligence, its ability to design purpose built crystalline materials remains in a nascent phase. At the moment inverse design is generally accomplished by either constraining the training data set or producing a vast number of samples from a generator network and constraining the output via post-processing. We show that a general adversarial network trained to produce crystal structures from a latent space can be fine tuned through the introduction of advanced graph neural networks as discriminators, including a universal force field, to intrinsically bias the network towards generation of target materials. This is exemplified utilizing two-dimensional topological insulators as a sample target space. While a number of two-dimensional topological insulators have been predicted, the size of the band-gap, a measure of topological protection, remains a concern in most candidate compounds.  The resulting generative network is shown to yield novel topological insulators. 
\end{abstract}

\maketitle
\par 
\section{Introduction}

The potential use of generative artificial intelligence to design crystalline materials satisfying a given set of electronic or structural properties has progressed rapidly in recent years\cite{sanchez2018inverse,xie2021crystal,long2021constrained, merchant2023scaling,zeni2023mattergen}. While current databases of organic and inorganic crystals contain millions of structures, this represents just a fraction of the material landscape\cite{jain2013commentary,saal2013materials,mounet2018two,borysov2017organic,haastrup2018computational}. Furthermore, high-throughput screenings for certain desirable properties have yet to reveal an optimal compound; a key example being large band-gap topological insulators\cite{vergniory2021all,tang2019efficient,zhang2019catalogue,bradlyn2019disconnected,2DSymmTopo,marrazzo2019relative,petralanda2024two}, the subject of this work. In the case of topological insulators, nearly all known inorganic crystals have been screened previously\cite{petralanda2024two,marrazzo2019relative,2DSymmTopo}, necessitating generation of novel compounds to fulfill the promise of topological materials in quantum technologies. 
\begin{figure}[h!]
    \centering
    \includegraphics[width=8cm]{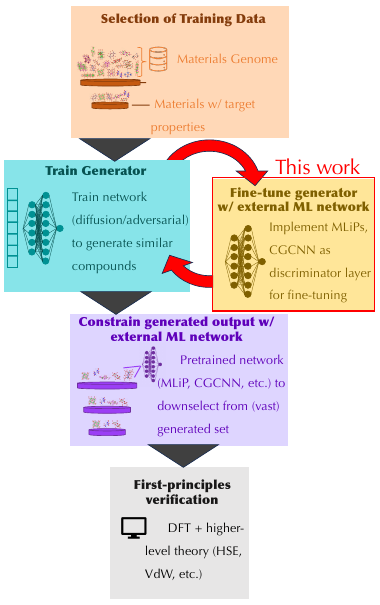}
    \caption{\textbf{Proposed deep generative workflow:} Standard workflow for deep generation of materials with target properties and proposed modification detailed in this work.}
    \label{fig:Fig0}
\end{figure}
\par 
Generative artificial intelligence (GAI) is well positioned to serve as a guide in exploring the material genome in manner which optimizes the odds of locating useful compounds. There exist multiple possible approaches to the development of GAI for prediction of crystal structures. One approach is that of Ref. \cite{merchant2023scaling}, in which chemical substitutions are made to existing compounds. For each unique chemical composition ab initio random structure searching is utilized to identify stable compounds. While this method has achieved immense success in expanding the number of crystal structures at or near the convex hull, the network is limited in its ability to generate crystal structures with target properties. 
\par 
An alternative approach to crystal structure generation is to implement a diffusion network\cite{xie2021crystal}. Stable diffusion networks for crystal structure generation can be crudely summarized by the following algorithm: (1) encode the crystal structures into a tensor, (2) introduce noise into the tensors providing them as input to a deep neural network with the target output being the clean, de-noised tensor, (3) once trained, tensors from a latent space can be supplied to the network with the output being a new crystal structure. The crystal diffusion variational autoencoder (CDVAE) in Ref. \cite{xie2021crystal} is one such network which has been widely implemented. 

\par 
In this work we pursue a third option, a generative adversarial network (GAN). GANs offer two main advantages over stable diffusion networks. First, by not relying on the diffusion process but rather closely mimicking a given target, the output of GANs is generally sharper. In the case of design for electronic properties this is vital as the same atomic composition in a different structural arrangement can lead to distinct electronic properties. Next, there is tremendous flexibility in construction of a discriminator layer for a GAN. Typically a GAN utilizes a single discriminator to distinguish real and fake samples, however in the context of crystal design, multiple discriminators can be utilized in parallel or serial to bias the generator network towards creation of stable, insulating and topological materials. This flexibility also allows for the power of universal machine learned interatomic potentials (MLiPs) to be incorporated in the training process\cite{batatia2022mace,deng2023chgnet,chen2022universal}. MLiPs have emerged as an incredibly powerful tool to efficiently predict forces and energies for a crystalline system without the requirement of expensive density functional theory computations. Importantly, this allows for relaxation and rapid estimation of phonon modes. 
\begin{figure}
    \centering
    \includegraphics[width=8cm]{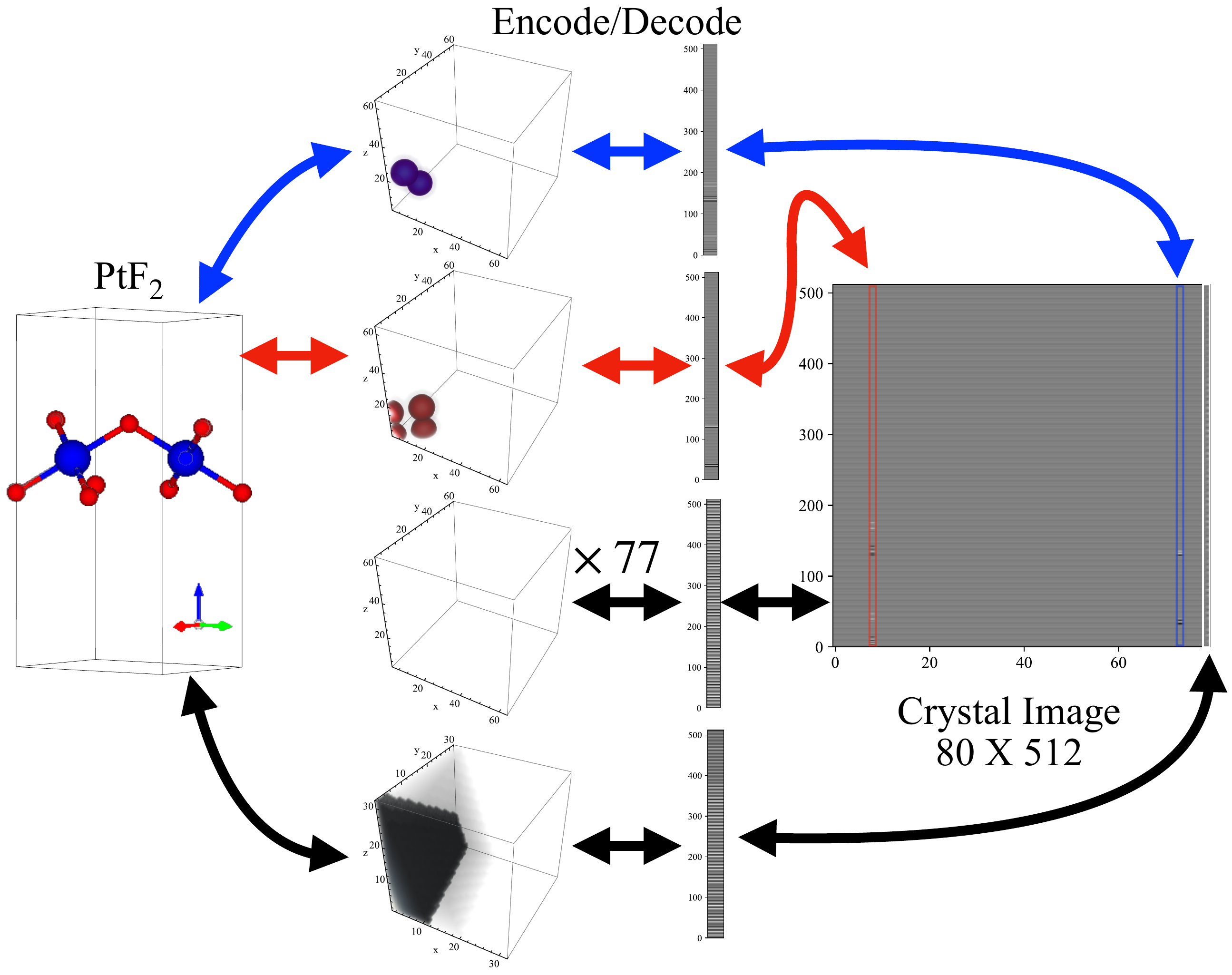}
    \caption{\textbf{Processing crystal structures for training:} Reversible process for generation of two-dimensional images representing crystal structures for use in training generative adversarial network via autoencoding of voxel images created for each constituent element and the unit cell.}
    \label{fig:ImageGen}
\end{figure}

\begin{figure*}
    \centering
    \includegraphics[width=18cm]{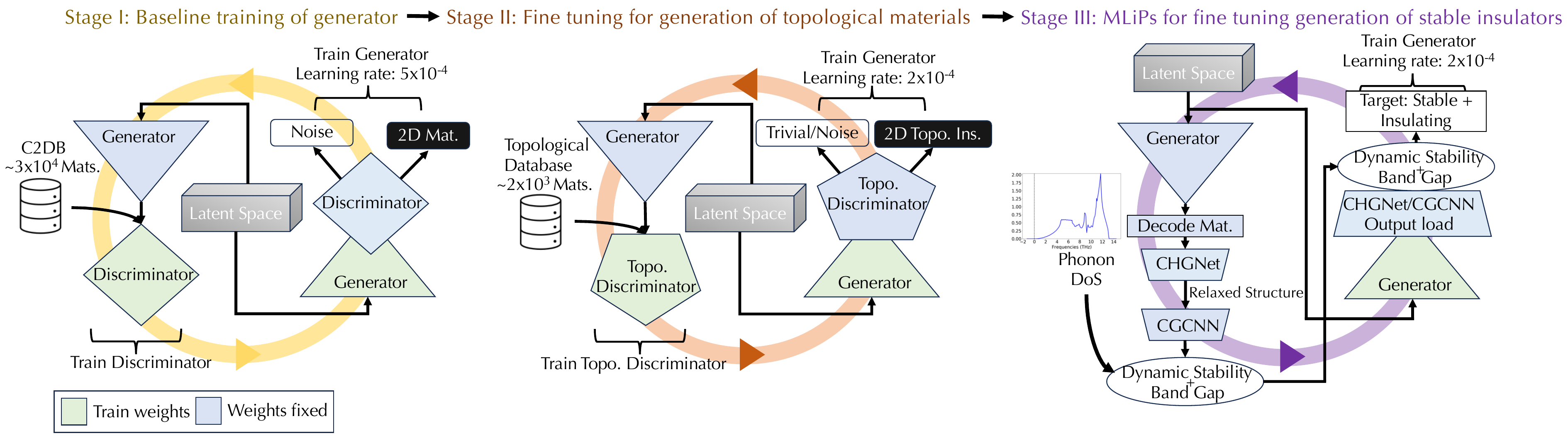}
    \caption{\textbf{Training generative network:} The three stage process for baseline training and fine-tuning of the generative network. Left: Baseline training of generative network for producing two-dimensional crystal structures via adversarial discriminator trained using two-dimensional crystal structures from the C2DB database\cite{haastrup2018computational}. Center: Fine-tuning the generative network by introduction of a new discriminator trained using the topological material dataset generated in Ref. \cite{TynerML}. Right: Final fine-tuning stage replacing the discriminator with a universal force-field for computation of the phonon density of states and a crystal graph convolutional network for bandgap prediction.}
    \label{fig:TrainProcess}
\end{figure*}
\par 
To leverage the power of MLiPs in the generative process, we construct a GAN for design of two-dimensional topological insulators. The GAN is unique in the manner with which fine tuning of the model parameters is accomplished. The baseline model is a familiar GAN architecture, trained with a single discriminator which distinguishes real from synthesized inputs representing crystal structures. A similar procedure has been implemented in Ref. \cite{long2021constrained}. Unlike Refs. \cite{long2021constrained,hong2025discovery}, we do not undergo a single stage of training, generate a vast number of potential compounds and then rely on auxiliary machine learning networks to constrain the output. This approach which has been established previously is represented in Fig. \eqref{fig:Fig0}. Rather, we supplement this approach with a careful fine-tuning procedure utilizing new discriminators. The first stage of fine-tuning considers a smaller dataset of 962 two-dimensional topological materials, the discriminator introduced in this stage distinguishes real samples supporting non-trivial topology from both topologically trivial real samples and generated (also denoted as synthetic) samples, which we presume to be topologically trivial. 
\par 
The second fine-tuning stage leverages the power of MLiPs, specifically CHGNet\cite{deng2023chgnet}. A custom discriminator layer calls this MLiP, relaxing the generated structure and computing the phonon density of states. Simultaneous to this evaluation by the MLiP, the band gap of the relaxed crystal structure is predicted via a pre-trained crystal graph convolutional neural network\cite{cgcnn,cgcnn_wolverton}. Though computationally expensive, this final layer leverages the most powerful machine learning networks in computational materials science. In doing so we are able to introduce bias directly into the generator rather than simply constraining the output. While the diversity of crystals generated upon training is reduced compared to stable diffusion networks, we provide evidence that this protocol increases the fraction of generated crystals with the target properties: stable, insulating and topological.

\par
\section{Training data and pre-processing}
\par 
We construct two distinct datasets for the GAN. The first dataset utilized to train the baseline GAN is the C2DB database\cite{haastrup2018computational}. This dataset will be used to train the a discriminator to simply distinguish between “real” crystals and “fake” crystal generated from the latent space. This database contains more than 15000 compounds, the largest available database of two-dimensional materials.  The second dataset contains two-dimensional topological materials. It is taken from Ref. \cite{TynerML}, in which two-dimensional insulators were screened to identify those that support a finite Chern or spin-Chern number\cite{TKNN1982,Niu1985,FuKane,Kane2005,Sheng2006,bernevig2006quantum,Prodan2009}. This dataset is advantageous as it does not discriminate between first-order topological insulators or higher-order/fragile topological insulators\cite{Benalcazar61,BenalacazarCn,schindler2018higher,Schindlereaat0346,song2020twisted,Fragile,tyner2020topology}. Previous catalogs of two-dimensional topological insulators had relied solely on symmetry indicator techniques to reveal a non-trivial Kane-Mele index or mirror Chern number. While efficient, these techniques label more exotic forms of topological insulators, such as higher-order systems, as trivial. In Ref. \cite{TynerML}, the criteria for being labeled as trivial was made more severe. To be labeled trivial, a magnetic flux tube inserted in the bulk must display no bound modes in the mid-gap. This is a strict criteria for a system being labeled trivial as it has been shown that a magnetic flux tube inserted in the bulk of all first order topological insulators will cause a spectral flow as the strength of the flux is tuned from $\phi=0$ to $\phi=\phi_{0}$ where $\phi_{0}=hc/e$\cite{QiSpinCharge,SpinChargeVishwanath,Wang_2010,slager2012,MESAROS2013977,tynerbismuthene,TynerRealSpace}. Similarly for higher-order or fragile insulators supporting a finite spin-Chern number as defined by Prodan\cite{Prodan2009}, but lacking gapless edge states, an inserted flux tube need not display spectral flow as flux is tuned, however, the flux tube will continue to bind a number of modes determined by the spin-Chern number.
\par 
The dataset in Ref. \cite{TynerML} contains 962 systems. While this number is reduced relative to alternative datasets, it remains preferable due to the lack of false negatives in the form of fragile and higher-order topological insulators. Furthermore, this dataset has shown prior success in predicting non-trivial topology and will be just one of three datasets used in training of the GAN. 

\par 
\subsection{Crystal structure representation}
The manner in which a crystal structure is represented for training a deep neural network is of vital importance. Crystalline compounds are inherently discrete objects, however convolutional neural networks are designed to work with objects that accommodate a continuous representation. We therefore pre-process all crystal structures in the training set such that they can be represented as continuous two-dimensional crystal images by autoencoding voxel images of the crystal structure to create a 2D crystal graph in a manner similar to that implemented in Ref. \cite{long2021constrained}. We account for the possibility of 79 different elements in the crystal structure, specifically atomic numbers 1-84 removing the noble gases. As a result, regardless of the number of elements in a single crystal structure, 80 voxel images will be produced. An autoencoder then translates each image into a vector. A similar process is done to form a voxel image of the lattice which again is translated into a one-dimensional array through use of an autoencoder. These one-dimensional arrays are then reshaped into 2D crystal images. In this way the crystal structure obtains a continuous and reversible representation. Details of the auto encoder are available in the supplementary information of Ref. \cite{TynerML}.

\par 
\section{Generative model design and training}
\par 
The architecture of the generative model employed in this work follows a widely implemented strategy of utilizing two-dimensional convolutional transpose layers to produce a tensor image from a vector in the latent space. The details of the network architecture are shown in Fig. \eqref{fig:TrainProcess}. The three part training process is detailed in the following sections. 
\par 
\subsection{Baseline training: Stage I}
\par
The first stage of training fixes the baseline model and is carried out with the intent to construct a generative model capable of outputting two-dimensional crystal structures. The C2DB database, containing more than 15000 structures, is utilized as the database of real-samples to train the binary discriminator shown on the left in Fig. \eqref{fig:TrainProcess} to distinguish synthesized (fake) crystal images from those corresponding to real crystal structures within the C2DB. At this stage we note that GANs are known to quickly overfit, mimicking a small set of the real data. To avoid this issue and ensure that the generative model outputs a diverse set of crystal structures, both in constituent element type and size of unit cell, the initial training is restricted to 100 epochs. At this point the generative model is saved and utilized to generate 1600 2D crystal structures. For these crystal structures both the atomic positions and lattice constants are fully relaxed within density functional theory as implemented via the Quantum Espresso software package\cite{Pizzi2020,Perdew1996}. The relaxation is done utilizing fully relativistic norm-conserving pseudo potentials from the pseudo dojo library\cite{van2018pseudodojo}, a grid of $20 \times 20 \times 1$ $\mathbf{k}$-points, plane wave cutoff of 80 Ry and spin-orbit coupling is included in each case. The output of this computation is then extracted and utilized to fine tune the CHGNet universal force field. 
\par 
\subsection{Fine tuning A: Stage II}
\par 
Having performed the baseline training to construct a generator capable of producing a diverse set of two-dimensional crystal structures, we now embark on the first stage of fine tuning. This involves replacing the discriminator layer of the GAN with one designed to perform binary topological classification of crystal images. As described previously, this binary classification is based on the presence/absence of a finite ground state spin-Chern number, thereby encompassing first-order, higher-order, and fragile topological phases. A similar model was constructed in Ref. \cite{TynerML} for the purposes of property prediction. 
\par 
The training protocol is nearly identical to that done in stage I however, when training the discriminator, targets for real crystal structures are not homogeneously labeled. Instead the targets follow their topological classification. Due to the reduced size of the topological material dataset, the learning rate is decreased from 0.0005 to 0.0002 and the GAN is trained for 50 epochs.  
\par 

\begin{figure}
    \centering
    \includegraphics[width=8cm]{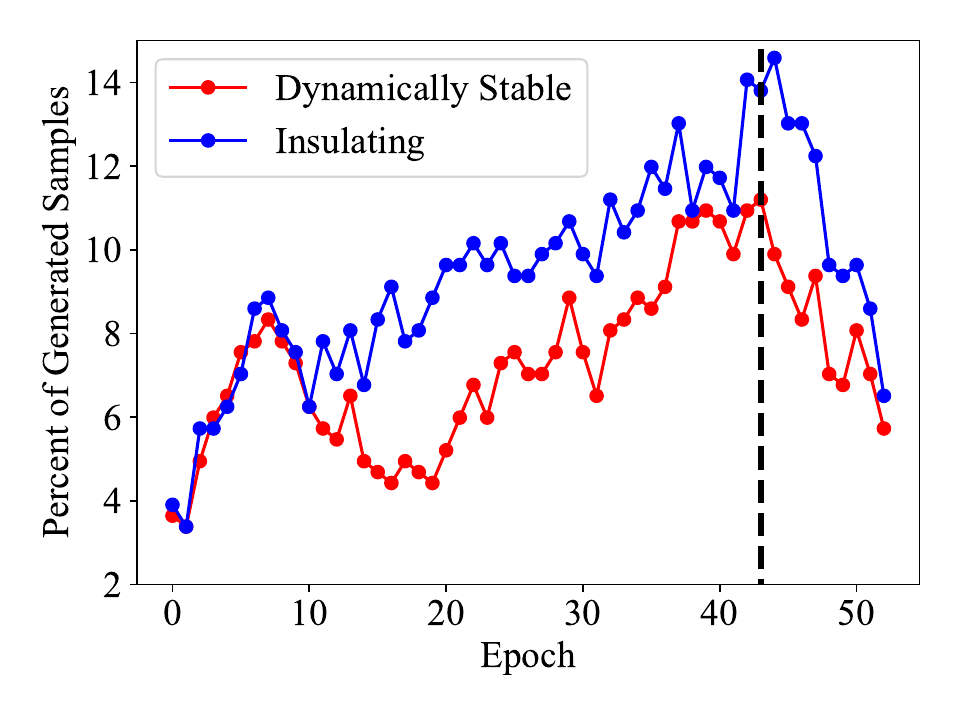}
    \caption{\textbf{Universal force field as a discriminator fine-tuning of generative network:} Percent of generated samples classified as dynamically stable via analysis of phonon modes by the universal force field and insulating by pre-trained crystal graph convolutional neural network in each epoch of the final fine-tuning stage.}
    \label{fig:MLPTrain}
\end{figure}
\subsection{Fine tuning B: Stage III}
\par 
Having fine tuned the model to bias towards generation of topological materials, we wish to perform further fine tuning in pursuit of stable and insulating two-dimensional materials. To do so we will take advantage of the flexible nature of GAN discriminators and the power of modern universal force fields. Namely, we fine-tune CHGNet\cite{deng2023chgnet} utilizing the data acquired in relaxation of the 1600 two-dimensional materials generated in stage I. Following this fine tuning, we construct a discriminator layer which takes as input the crystal image. It then decodes the image to extract the crystal structure. This crystal structure is relaxed using the structure relaxer function within CHGNet. The phonon density of states for the relaxed structure is then computed using CHGNet. The number of negative frequency modes is then quantified. The discriminator outputs a value between zero and one based on a high density of negative phonon modes (zero) or absence of negative phonon modes (one). As the target for the GAN is ones, the weights are fine-tuned to bias towards dynamic stability.

\begin{figure*}
    \centering
    \includegraphics[width=18cm]{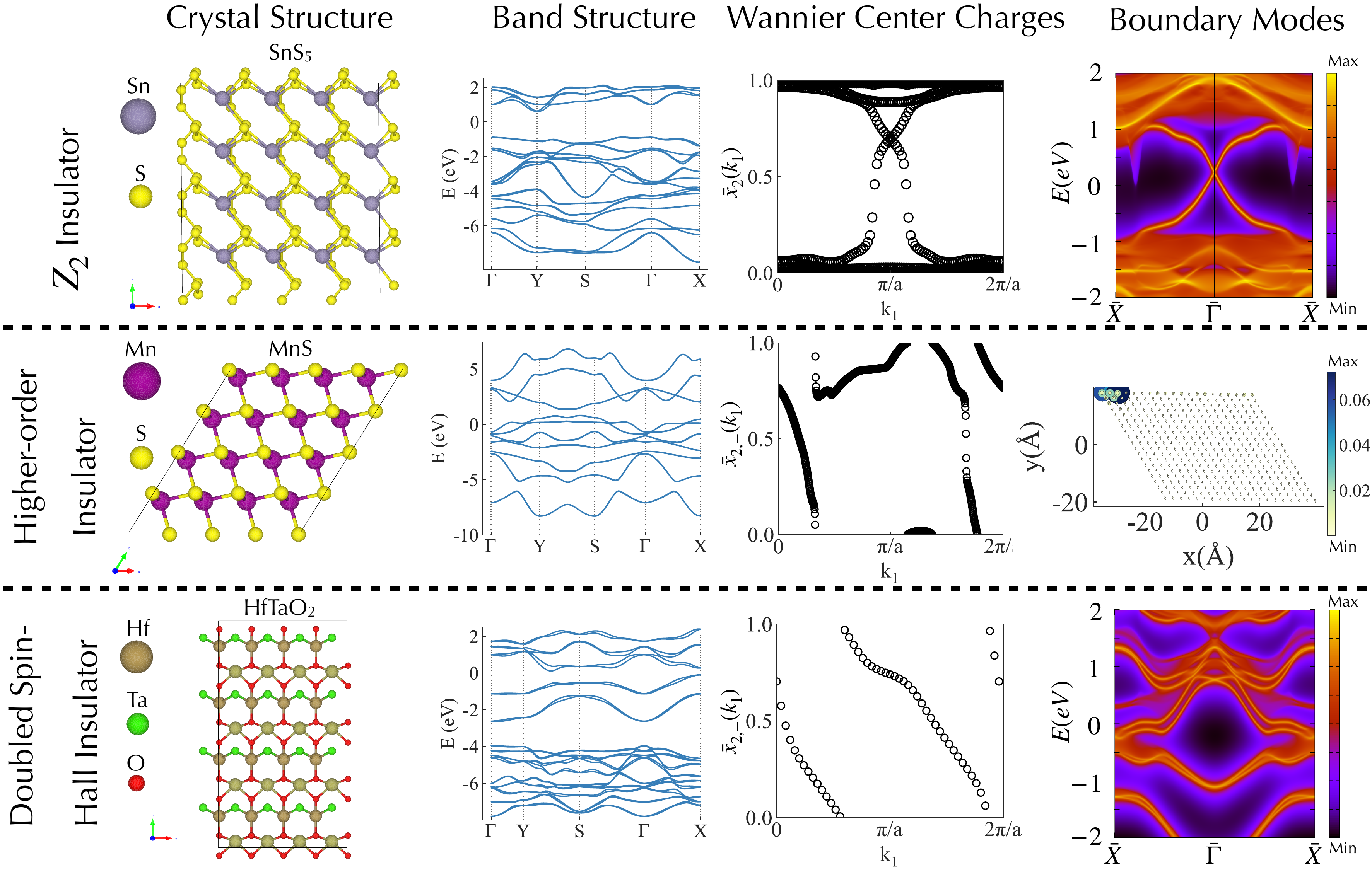}
    \caption{\textbf{Generated topological materials: } The crystal structure and band structure of three generated topological materials is shown after undergoing relaxation within density functional theory in the first two columns. The bulk topology is analyzed by inspection of the Wannier center charges to determine the non-trivial $Z_{2}$ index for SnS$_{5}$ and spin-resolved Wannier center charges for determining the spin-Chern number $|C_{s}|=2$ for MnS and HfTaO$_{2}$. The final column details evidence of topological boundary modes. For SnS$_{5}$ plotting the (01) surface spectral density reveals a single set of gapless helical edge states. For MnS, examining the spatial localization of mid-gap modes for a slab reveals corner localized states. In HfTaO$_{2}$, examining the (01) surface spectral density shows two sets of helical edge states intersecting the Fermi energy.}
    \label{fig:mats}
\end{figure*}
\par 
Finally to further bias towards generation of insulators, a second discriminator taking the form of a convolutional crystal graph network (CGCNN), is used to predict the band gap of the crystal structure relaxed by CHGNet. The CGCNN is trained previously using the C2DB database with details available in the supplementary material. Acting as a discriminator, the output of the CGCNN is scaled to fall between zero and one, zero corresponding to a metal and one corresponding to an insulator with a band gap above 1$eV$. This is schematically outlined in the cycle on the right in Fig. \eqref{fig:TrainProcess}
\par 
Unlike the previous stages, real-samples are not necessary to train the discriminator in this stage as the constituent components of the discriminator, CGCNN and CHGNet, are extensively pre-trained. We can therefore simply train the GAN for an additional 50 cycles. The accuracy per epoch in this stage of training corresponds to the percentage of generated compounds determined to be stable and insulating by the CHGNet force-field computation of the phonon density of states and trained CGCNN model band-gap prediction respectively. In Fig. \eqref{fig:MLPTrain} the model accuracy per epoch is shown to increase progressively reaching a maximum of 14.2\% insulating and 11.3\% dynamically stable at epoch 43 at which point early-stopping is implemented. 
\par 
It is important to emphasize that the success rate for deep generative networks in producing crystalline materials with target properties is a current issue. In a recent work, Ref. \cite{hong2025discovery}, from 10000 materials generated using the CDVAE diffusion network trained on topological systems, four topological insulators and sixteen topological semimetals were identified corresponding to a success rate of 0.2\%. This does not detract from the importance or novelty of the work; rather this situation underscores the immense challenge in successfully generating target materials. Moreover, in Ref. \cite{hong2025discovery} \emph{no} topological insulators with a direct band gap were identified. This is understandable as genuine topological insulators generally support minimal bad-gaps generated by inclusion of spin-orbit coupling. 
\par

With this perspective on the state of the field in mind, we note that final fine-tuning cycle has a remarkable effect of potentially raising the success rate by $>400\%$; putting the power of modern universal force fields on full display. To truly verify this enhanced performance we must also perform DFT validation on a set of generated compounds. To do so, 200 crystal structures are then generated. We purposefully choose to generate a relatively small number of candidates, as opposed to the 10000 or more generated in other works, as the purpose of this modified training program is to create generative models with an increased success rate. In doing so we provide a route to alleviate the need to generate thousands of synthetic crystal structures to identify optimal candidates with the desired target properties.

\section{Analysis of generated materials}
\par 
The materials generated by the finalized generator are first relaxed within DFT such that interatomic forces are below $10^{-5} Ry/Bohr$. The relaxed structures are then examined with insulators being isolated to screen for non-trivial topology. Topological screening involves creation of a Wannier tight-binding model using the Wannier90 software package. The $Z_{2}$ index, along with the spin-Chern number is computed utilizing the BerryEasy software packages\cite{tyner2023berryeasy}. Edge states are further examined using the WannierTools software package\cite{WU2017}. The crystal structure of all confirmed topological materials is available on GitHub\cite{github} and listed in the supplementary material. 
\par 
Of the 200 materials screened we identify 7 insulators displaying evidence of non-trivial topology, corresponding to a comparatively high success rate of 3.5\%. In this section we examine three generated compounds falling into the categories of (I) $Z_{2}$ topological insulator, (II) higher-order topological insulator, and (III) doubled spin-Hall insulator. 
\par 
\emph{$\mathcal{Z}_{2}$ topological insulator:} Two-dimensional $Z_{2}$ topological insulators are of wide interest to the community for their potential application in numerous quantum technologies including quantum sensor and topological quantum computers. Perhaps the most famous class of two-dimensional insulators with a non-trivial Fu-Kane $Z_{2}$ classification are the $1T'$ transition metal dichalcogenides $MX_{2}$ where M=(Mo,W) and X=(Te,S,Se)\cite{qian2014quantum}. The signature of this state of matter is an insulating bulk and gapless helical surface states. The bulk topology can further be diagnosed by the existence of a gapless Wannier center charge spectra as detailed in Refs. \cite{yu2011equivalent,Z2pack,Taherinejad2014}.
\par 
As discussed previously, despite the attractive properties of $Z_{2}$ topological insulators, a common bottleneck to their experimental utility is the extremely small size of the bulk band gap, a measure of the stability of the topological phase. The current record band-gap for a predicted $Z_{2}$ topological insulator is  $1.34 eV$, associated with plumbene films with chemical decoration, PbX (X=H, F, Cl, Br and I) monolayers\cite{zhao2016unexpected}. 
\par 
Here we present SnS$_{5}$ as a candidate $Z_{2}$ topological insulator generated by the trained neural network. The crystal and electronic band structure are shown in the top row of Fig. \eqref{fig:mats}. We note the band-gap is of a magnitude $1.49eV$, placing it in contention with the largest gap $Z_{2}$ topological insulators. For details of phonon modes please consult the appendix. The gapless Wannier center charge spectra confirming the $Z_{2}$ classification is shown in Fig. \eqref{fig:mats} along with the presence of gappless helical edge states on the (01) surface.

\par 
\emph{Higher-order topological insulator:} 
Higher-order topology in two-dimensional materials is identified by the presence of bulk and surface electronic gaps, while corner-localized mid-gap states are present in when open boundary conditions are applied along both principal directions\cite{schindler2018higher,Schindlereaat0346,Benalcazar61,BenalacazarCn}. Higher-order topology has been studied at length in recent years due to its prevalence in many systems of interest such as $1H$ transmission metal dichalcogenides, twisted bilayer graphene, and more. The bulk topology of higher-order topological insulators has been associated with the presence of a non-zero spin-Chern number\cite{costa2021discovery}, $C_{s}=(C_{\uparrow}-C_{\downarrow})/2=2$, computed using the formalism of Prodan\cite{Prodan2009}, and recently developed further by Lin et. al\cite{Lin2022Spin}. Here we put forth MnS, generated by the trained network, as a predicted higher-order topological insulator. The crystal structure and band structure are shown in the middle row of Fig. \eqref{fig:mats} details that the system is insulating. The spin-resolved Wannier-center charges, computed for the preferred spin-direction, $\hat{\mathbf{n}}=\sin\phi\sigma_{z}+\cos\phi\sigma_{x}$ with $\phi=\pi/8$, are shown detailing $C_{s}=2$. This computation suggests the presence of non-trivial higher-order topology which is further confirmed by analyzing the localization of eigenstates in the mid-gap for $20 \times 20$ unit cell slab with open boundary conditions. Among the mid-gap states we identify corner-localized modes with the spatial localization of one such mode shown on the right side of the second row in Fig. \eqref{fig:mats}.

\par 
\emph{Doubled spin-Hall insulator:} 
Before concluding, we discuss a third material of interest, HTaClO$_{2}$, put forth by the generative network which is shown to support a spin-Chern number $C_{s}=2$ as associated with $Z_{2}$ trivial topological phases such as higher-order insulators, however, in this case we find that the bulk topology gives rise to two-sets of helical edge states rather than the single set associated with a $Z_{2}$ insulator. Spin-Hall insulators supporting a doubled spin-Chern number and two sets of helical edge states have been previously predicted in twisted WSe$_{2}$\cite{kang2024double} and $\alpha$-antimonene as well as $\alpha$-bismuthene\cite{Antimonene}. These phase are of interest to the theoretical and experimental community due to their enhanced transport response, namely an enhanced spin-Hall conductivity. However, as these systems are $Z_{2}$ trivial the gapless edge states can be be unstable and gapped be realistic considerations such as growth on a substrate, causing a phase transition to a higher-order topological state. Nonetheless, such systems have attracted wide interest in recent years for going beyond the $Z_{2}$ paradigm. It is thus of note that the generator has produced such a topological phase. 

\section{Discussion}
In this work we explore whether the flexible architecture  of GANs allows for incorporation of multiple stages of fine tuning in the training process to bias the network towards creation of specialized materials. This is especially useful for inverse design of materials for which the desired properties are not known for a large dataset. Rather, we show that a baseline GAN can be successfully fine-tuned using a smaller dataset to accomplish a more precise form of inverse design. We expect this architecture and strategy to therefore find great use in a wide range of materials design efforts. 
\par

\acknowledgments{}
The computations were enabled by resources provided by the National Academic Infrastructure for Supercomputing in Sweden (NAISS), partially funded by the Swedish Research Council through grant agreement no. 2022-06725. NORDITA is supported in part by NordForsk.

\bibliographystyle{apsrev4-1}
\nocite{apsrev41Control}
\bibliography{Ref.bib}

\begin{thebibliography}{60}%
\makeatletter
\providecommand \@ifxundefined [1]{%
 \@ifx{#1\undefined}
}%
\providecommand \@ifnum [1]{%
 \ifnum #1\expandafter \@firstoftwo
 \else \expandafter \@secondoftwo
 \fi
}%
\providecommand \@ifx [1]{%
 \ifx #1\expandafter \@firstoftwo
 \else \expandafter \@secondoftwo
 \fi
}%
\providecommand \natexlab [1]{#1}%
\providecommand \enquote  [1]{``#1''}%
\providecommand \bibnamefont  [1]{#1}%
\providecommand \bibfnamefont [1]{#1}%
\providecommand \citenamefont [1]{#1}%
\providecommand \href@noop [0]{\@secondoftwo}%
\providecommand \href [0]{\begingroup \@sanitize@url \@href}%
\providecommand \@href[1]{\@@startlink{#1}\@@href}%
\providecommand \@@href[1]{\endgroup#1\@@endlink}%
\providecommand \@sanitize@url [0]{\catcode `\\12\catcode `\$12\catcode `\&12\catcode `\#12\catcode `\^12\catcode `\_12\catcode `\%12\relax}%
\providecommand \@@startlink[1]{}%
\providecommand \@@endlink[0]{}%
\providecommand \url  [0]{\begingroup\@sanitize@url \@url }%
\providecommand \@url [1]{\endgroup\@href {#1}{\urlprefix }}%
\providecommand \urlprefix  [0]{URL }%
\providecommand \Eprint [0]{\href }%
\providecommand \doibase [0]{http://dx.doi.org/}%
\providecommand \selectlanguage [0]{\@gobble}%
\providecommand \bibinfo  [0]{\@secondoftwo}%
\providecommand \bibfield  [0]{\@secondoftwo}%
\providecommand \translation [1]{[#1]}%
\providecommand \BibitemOpen [0]{}%
\providecommand \bibitemStop [0]{}%
\providecommand \bibitemNoStop [0]{.\EOS\space}%
\providecommand \EOS [0]{\spacefactor3000\relax}%
\providecommand \BibitemShut  [1]{\csname bibitem#1\endcsname}%
\let\auto@bib@innerbib\@empty
\bibitem [{\citenamefont {Sanchez-Lengeling}\ and\ \citenamefont {Aspuru-Guzik}(2018)}]{sanchez2018inverse}%
  \BibitemOpen
  \bibfield  {author} {\bibinfo {author} {\bibfnamefont {B.}~\bibnamefont {Sanchez-Lengeling}}\ and\ \bibinfo {author} {\bibfnamefont {A.}~\bibnamefont {Aspuru-Guzik}},\ }\href {\doibase 10.1126/science.aat2663} {\bibfield  {journal} {\bibinfo  {journal} {Science}\ }\textbf {\bibinfo {volume} {361}},\ \bibinfo {pages} {360} (\bibinfo {year} {2018})}\BibitemShut {NoStop}%
\bibitem [{\citenamefont {Xie}\ \emph {et~al.}(2021)\citenamefont {Xie}, \citenamefont {Fu}, \citenamefont {Ganea}, \citenamefont {Barzilay},\ and\ \citenamefont {Jaakkola}}]{xie2021crystal}%
  \BibitemOpen
  \bibfield  {author} {\bibinfo {author} {\bibfnamefont {T.}~\bibnamefont {Xie}}, \bibinfo {author} {\bibfnamefont {X.}~\bibnamefont {Fu}}, \bibinfo {author} {\bibfnamefont {O.-E.}\ \bibnamefont {Ganea}}, \bibinfo {author} {\bibfnamefont {R.}~\bibnamefont {Barzilay}}, \ and\ \bibinfo {author} {\bibfnamefont {T.}~\bibnamefont {Jaakkola}},\ }\href@noop {} {\bibfield  {journal} {\bibinfo  {journal} {arXiv:2110.06197}\ } (\bibinfo {year} {2021})}\BibitemShut {NoStop}%
\bibitem [{\citenamefont {Long}\ \emph {et~al.}(2021)\citenamefont {Long}, \citenamefont {Fortunato}, \citenamefont {Opahle}, \citenamefont {Zhang}, \citenamefont {Samathrakis}, \citenamefont {Shen}, \citenamefont {Gutfleisch},\ and\ \citenamefont {Zhang}}]{long2021constrained}%
  \BibitemOpen
  \bibfield  {author} {\bibinfo {author} {\bibfnamefont {T.}~\bibnamefont {Long}}, \bibinfo {author} {\bibfnamefont {N.~M.}\ \bibnamefont {Fortunato}}, \bibinfo {author} {\bibfnamefont {I.}~\bibnamefont {Opahle}}, \bibinfo {author} {\bibfnamefont {Y.}~\bibnamefont {Zhang}}, \bibinfo {author} {\bibfnamefont {I.}~\bibnamefont {Samathrakis}}, \bibinfo {author} {\bibfnamefont {C.}~\bibnamefont {Shen}}, \bibinfo {author} {\bibfnamefont {O.}~\bibnamefont {Gutfleisch}}, \ and\ \bibinfo {author} {\bibfnamefont {H.}~\bibnamefont {Zhang}},\ }\href {\doibase https://doi.org/10.1038/s41524-021-00526-4} {\bibfield  {journal} {\bibinfo  {journal} {npj Computational Materials}\ }\textbf {\bibinfo {volume} {7}},\ \bibinfo {pages} {66} (\bibinfo {year} {2021})}\BibitemShut {NoStop}%
\bibitem [{\citenamefont {Merchant}\ \emph {et~al.}(2023)\citenamefont {Merchant}, \citenamefont {Batzner}, \citenamefont {Schoenholz}, \citenamefont {Aykol}, \citenamefont {Cheon},\ and\ \citenamefont {Cubuk}}]{merchant2023scaling}%
  \BibitemOpen
  \bibfield  {author} {\bibinfo {author} {\bibfnamefont {A.}~\bibnamefont {Merchant}}, \bibinfo {author} {\bibfnamefont {S.}~\bibnamefont {Batzner}}, \bibinfo {author} {\bibfnamefont {S.~S.}\ \bibnamefont {Schoenholz}}, \bibinfo {author} {\bibfnamefont {M.}~\bibnamefont {Aykol}}, \bibinfo {author} {\bibfnamefont {G.}~\bibnamefont {Cheon}}, \ and\ \bibinfo {author} {\bibfnamefont {E.~D.}\ \bibnamefont {Cubuk}},\ }\href {\doibase https://doi.org/10.1038/s41586-023-06735-9} {\bibfield  {journal} {\bibinfo  {journal} {Nature}\ }\textbf {\bibinfo {volume} {624}},\ \bibinfo {pages} {80} (\bibinfo {year} {2023})}\BibitemShut {NoStop}%
\bibitem [{\citenamefont {Zeni}\ \emph {et~al.}(2023)\citenamefont {Zeni}, \citenamefont {Pinsler}, \citenamefont {Z{\"u}gner}, \citenamefont {Fowler}, \citenamefont {Horton}, \citenamefont {Fu}, \citenamefont {Shysheya}, \citenamefont {Crabb{\'e}}, \citenamefont {Sun}, \citenamefont {Smith} \emph {et~al.}}]{zeni2023mattergen}%
  \BibitemOpen
  \bibfield  {author} {\bibinfo {author} {\bibfnamefont {C.}~\bibnamefont {Zeni}}, \bibinfo {author} {\bibfnamefont {R.}~\bibnamefont {Pinsler}}, \bibinfo {author} {\bibfnamefont {D.}~\bibnamefont {Z{\"u}gner}}, \bibinfo {author} {\bibfnamefont {A.}~\bibnamefont {Fowler}}, \bibinfo {author} {\bibfnamefont {M.}~\bibnamefont {Horton}}, \bibinfo {author} {\bibfnamefont {X.}~\bibnamefont {Fu}}, \bibinfo {author} {\bibfnamefont {S.}~\bibnamefont {Shysheya}}, \bibinfo {author} {\bibfnamefont {J.}~\bibnamefont {Crabb{\'e}}}, \bibinfo {author} {\bibfnamefont {L.}~\bibnamefont {Sun}}, \bibinfo {author} {\bibfnamefont {J.}~\bibnamefont {Smith}},  \emph {et~al.},\ }\href@noop {} {\bibfield  {journal} {\bibinfo  {journal} {arXiv preprint arXiv:2312.03687}\ } (\bibinfo {year} {2023})}\BibitemShut {NoStop}%
\bibitem [{\citenamefont {Jain}\ \emph {et~al.}(2013)\citenamefont {Jain}, \citenamefont {Ong}, \citenamefont {Hautier}, \citenamefont {Chen}, \citenamefont {Richards}, \citenamefont {Dacek}, \citenamefont {Cholia}, \citenamefont {Gunter}, \citenamefont {Skinner}, \citenamefont {Ceder} \emph {et~al.}}]{jain2013commentary}%
  \BibitemOpen
  \bibfield  {author} {\bibinfo {author} {\bibfnamefont {A.}~\bibnamefont {Jain}}, \bibinfo {author} {\bibfnamefont {S.~P.}\ \bibnamefont {Ong}}, \bibinfo {author} {\bibfnamefont {G.}~\bibnamefont {Hautier}}, \bibinfo {author} {\bibfnamefont {W.}~\bibnamefont {Chen}}, \bibinfo {author} {\bibfnamefont {W.~D.}\ \bibnamefont {Richards}}, \bibinfo {author} {\bibfnamefont {S.}~\bibnamefont {Dacek}}, \bibinfo {author} {\bibfnamefont {S.}~\bibnamefont {Cholia}}, \bibinfo {author} {\bibfnamefont {D.}~\bibnamefont {Gunter}}, \bibinfo {author} {\bibfnamefont {D.}~\bibnamefont {Skinner}}, \bibinfo {author} {\bibfnamefont {G.}~\bibnamefont {Ceder}},  \emph {et~al.},\ }\href {\doibase https://doi.org/10.1063/1.4812323} {\bibfield  {journal} {\bibinfo  {journal} {APL materials}\ }\textbf {\bibinfo {volume} {1}} (\bibinfo {year} {2013}),\ https://doi.org/10.1063/1.4812323}\BibitemShut {NoStop}%
\bibitem [{\citenamefont {Saal}\ \emph {et~al.}(2013)\citenamefont {Saal}, \citenamefont {Kirklin}, \citenamefont {Aykol}, \citenamefont {Meredig},\ and\ \citenamefont {Wolverton}}]{saal2013materials}%
  \BibitemOpen
  \bibfield  {author} {\bibinfo {author} {\bibfnamefont {J.~E.}\ \bibnamefont {Saal}}, \bibinfo {author} {\bibfnamefont {S.}~\bibnamefont {Kirklin}}, \bibinfo {author} {\bibfnamefont {M.}~\bibnamefont {Aykol}}, \bibinfo {author} {\bibfnamefont {B.}~\bibnamefont {Meredig}}, \ and\ \bibinfo {author} {\bibfnamefont {C.}~\bibnamefont {Wolverton}},\ }\href {\doibase https://doi.org/10.1007/s11837-013-0755-4} {\bibfield  {journal} {\bibinfo  {journal} {Jom}\ }\textbf {\bibinfo {volume} {65}},\ \bibinfo {pages} {1501} (\bibinfo {year} {2013})}\BibitemShut {NoStop}%
\bibitem [{\citenamefont {Mounet}\ \emph {et~al.}(2018)\citenamefont {Mounet}, \citenamefont {Gibertini}, \citenamefont {Schwaller}, \citenamefont {Campi}, \citenamefont {Merkys}, \citenamefont {Marrazzo}, \citenamefont {Sohier}, \citenamefont {Castelli}, \citenamefont {Cepellotti}, \citenamefont {Pizzi} \emph {et~al.}}]{mounet2018two}%
  \BibitemOpen
  \bibfield  {author} {\bibinfo {author} {\bibfnamefont {N.}~\bibnamefont {Mounet}}, \bibinfo {author} {\bibfnamefont {M.}~\bibnamefont {Gibertini}}, \bibinfo {author} {\bibfnamefont {P.}~\bibnamefont {Schwaller}}, \bibinfo {author} {\bibfnamefont {D.}~\bibnamefont {Campi}}, \bibinfo {author} {\bibfnamefont {A.}~\bibnamefont {Merkys}}, \bibinfo {author} {\bibfnamefont {A.}~\bibnamefont {Marrazzo}}, \bibinfo {author} {\bibfnamefont {T.}~\bibnamefont {Sohier}}, \bibinfo {author} {\bibfnamefont {I.~E.}\ \bibnamefont {Castelli}}, \bibinfo {author} {\bibfnamefont {A.}~\bibnamefont {Cepellotti}}, \bibinfo {author} {\bibfnamefont {G.}~\bibnamefont {Pizzi}},  \emph {et~al.},\ }\href {\doibase https://doi.org/10.1038/s41565-017-0035-5} {\bibfield  {journal} {\bibinfo  {journal} {Nature nanotechnology}\ }\textbf {\bibinfo {volume} {13}},\ \bibinfo {pages} {246} (\bibinfo {year} {2018})}\BibitemShut {NoStop}%
\bibitem [{\citenamefont {Borysov}\ \emph {et~al.}(2017)\citenamefont {Borysov}, \citenamefont {Geilhufe},\ and\ \citenamefont {Balatsky}}]{borysov2017organic}%
  \BibitemOpen
  \bibfield  {author} {\bibinfo {author} {\bibfnamefont {S.~S.}\ \bibnamefont {Borysov}}, \bibinfo {author} {\bibfnamefont {R.~M.}\ \bibnamefont {Geilhufe}}, \ and\ \bibinfo {author} {\bibfnamefont {A.~V.}\ \bibnamefont {Balatsky}},\ }\href {\doibase https://doi.org/10.1371/journal.pone.0171501} {\bibfield  {journal} {\bibinfo  {journal} {PloS one}\ }\textbf {\bibinfo {volume} {12}},\ \bibinfo {pages} {e0171501} (\bibinfo {year} {2017})}\BibitemShut {NoStop}%
\bibitem [{\citenamefont {Haastrup}\ \emph {et~al.}(2018)\citenamefont {Haastrup}, \citenamefont {Strange}, \citenamefont {Pandey}, \citenamefont {Deilmann}, \citenamefont {Schmidt}, \citenamefont {Hinsche}, \citenamefont {Gjerding}, \citenamefont {Torelli}, \citenamefont {Larsen}, \citenamefont {Riis-Jensen} \emph {et~al.}}]{haastrup2018computational}%
  \BibitemOpen
  \bibfield  {author} {\bibinfo {author} {\bibfnamefont {S.}~\bibnamefont {Haastrup}}, \bibinfo {author} {\bibfnamefont {M.}~\bibnamefont {Strange}}, \bibinfo {author} {\bibfnamefont {M.}~\bibnamefont {Pandey}}, \bibinfo {author} {\bibfnamefont {T.}~\bibnamefont {Deilmann}}, \bibinfo {author} {\bibfnamefont {P.~S.}\ \bibnamefont {Schmidt}}, \bibinfo {author} {\bibfnamefont {N.~F.}\ \bibnamefont {Hinsche}}, \bibinfo {author} {\bibfnamefont {M.~N.}\ \bibnamefont {Gjerding}}, \bibinfo {author} {\bibfnamefont {D.}~\bibnamefont {Torelli}}, \bibinfo {author} {\bibfnamefont {P.~M.}\ \bibnamefont {Larsen}}, \bibinfo {author} {\bibfnamefont {A.~C.}\ \bibnamefont {Riis-Jensen}},  \emph {et~al.},\ }\href {\doibase 10.1088/2053-1583/aacfc1} {\bibfield  {journal} {\bibinfo  {journal} {2D Materials}\ }\textbf {\bibinfo {volume} {5}},\ \bibinfo {pages} {042002} (\bibinfo {year} {2018})}\BibitemShut {NoStop}%
\bibitem [{\citenamefont {Vergniory}\ \emph {et~al.}(2021)\citenamefont {Vergniory}, \citenamefont {Wieder}, \citenamefont {Elcoro}, \citenamefont {Parkin}, \citenamefont {Felser}, \citenamefont {Bernevig},\ and\ \citenamefont {Regnault}}]{vergniory2021all}%
  \BibitemOpen
  \bibfield  {author} {\bibinfo {author} {\bibfnamefont {M.~G.}\ \bibnamefont {Vergniory}}, \bibinfo {author} {\bibfnamefont {B.~J.}\ \bibnamefont {Wieder}}, \bibinfo {author} {\bibfnamefont {L.}~\bibnamefont {Elcoro}}, \bibinfo {author} {\bibfnamefont {S.~S.}\ \bibnamefont {Parkin}}, \bibinfo {author} {\bibfnamefont {C.}~\bibnamefont {Felser}}, \bibinfo {author} {\bibfnamefont {B.~A.}\ \bibnamefont {Bernevig}}, \ and\ \bibinfo {author} {\bibfnamefont {N.}~\bibnamefont {Regnault}},\ }\href@noop {} {\bibfield  {journal} {\bibinfo  {journal} {arXiv:2105.09954}\ } (\bibinfo {year} {2021})}\BibitemShut {NoStop}%
\bibitem [{\citenamefont {Tang}\ \emph {et~al.}(2019)\citenamefont {Tang}, \citenamefont {Po}, \citenamefont {Vishwanath},\ and\ \citenamefont {Wan}}]{tang2019efficient}%
  \BibitemOpen
  \bibfield  {author} {\bibinfo {author} {\bibfnamefont {F.}~\bibnamefont {Tang}}, \bibinfo {author} {\bibfnamefont {H.~C.}\ \bibnamefont {Po}}, \bibinfo {author} {\bibfnamefont {A.}~\bibnamefont {Vishwanath}}, \ and\ \bibinfo {author} {\bibfnamefont {X.}~\bibnamefont {Wan}},\ }\href {\doibase 10.1038/s41586-019-0937-5} {\bibfield  {journal} {\bibinfo  {journal} {Nat. Phys.}\ }\textbf {\bibinfo {volume} {15}},\ \bibinfo {pages} {470} (\bibinfo {year} {2019})}\BibitemShut {NoStop}%
\bibitem [{\citenamefont {Zhang}\ \emph {et~al.}(2019)\citenamefont {Zhang}, \citenamefont {Jiang}, \citenamefont {Song}, \citenamefont {Huang}, \citenamefont {He}, \citenamefont {Fang}, \citenamefont {Weng},\ and\ \citenamefont {Fang}}]{zhang2019catalogue}%
  \BibitemOpen
  \bibfield  {author} {\bibinfo {author} {\bibfnamefont {T.}~\bibnamefont {Zhang}}, \bibinfo {author} {\bibfnamefont {Y.}~\bibnamefont {Jiang}}, \bibinfo {author} {\bibfnamefont {Z.}~\bibnamefont {Song}}, \bibinfo {author} {\bibfnamefont {H.}~\bibnamefont {Huang}}, \bibinfo {author} {\bibfnamefont {Y.}~\bibnamefont {He}}, \bibinfo {author} {\bibfnamefont {Z.}~\bibnamefont {Fang}}, \bibinfo {author} {\bibfnamefont {H.}~\bibnamefont {Weng}}, \ and\ \bibinfo {author} {\bibfnamefont {C.}~\bibnamefont {Fang}},\ }\href {\doibase 10.1038/s41586-019-0944-6} {\bibfield  {journal} {\bibinfo  {journal} {Nature}\ }\textbf {\bibinfo {volume} {566}},\ \bibinfo {pages} {475} (\bibinfo {year} {2019})}\BibitemShut {NoStop}%
\bibitem [{\citenamefont {Bradlyn}\ \emph {et~al.}(2019)\citenamefont {Bradlyn}, \citenamefont {Wang}, \citenamefont {Cano},\ and\ \citenamefont {Bernevig}}]{bradlyn2019disconnected}%
  \BibitemOpen
  \bibfield  {author} {\bibinfo {author} {\bibfnamefont {B.}~\bibnamefont {Bradlyn}}, \bibinfo {author} {\bibfnamefont {Z.}~\bibnamefont {Wang}}, \bibinfo {author} {\bibfnamefont {J.}~\bibnamefont {Cano}}, \ and\ \bibinfo {author} {\bibfnamefont {B.~A.}\ \bibnamefont {Bernevig}},\ }\href {\doibase 10.1103/PhysRevB.99.045140} {\bibfield  {journal} {\bibinfo  {journal} {Phys. Rev. B}\ }\textbf {\bibinfo {volume} {99}},\ \bibinfo {pages} {045140} (\bibinfo {year} {2019})}\BibitemShut {NoStop}%
\bibitem [{\citenamefont {Wang}\ \emph {et~al.}(2019)\citenamefont {Wang}, \citenamefont {Tang}, \citenamefont {Ji}, \citenamefont {Zhang}, \citenamefont {Vishwanath}, \citenamefont {Po},\ and\ \citenamefont {Wan}}]{2DSymmTopo}%
  \BibitemOpen
  \bibfield  {author} {\bibinfo {author} {\bibfnamefont {D.}~\bibnamefont {Wang}}, \bibinfo {author} {\bibfnamefont {F.}~\bibnamefont {Tang}}, \bibinfo {author} {\bibfnamefont {J.}~\bibnamefont {Ji}}, \bibinfo {author} {\bibfnamefont {W.}~\bibnamefont {Zhang}}, \bibinfo {author} {\bibfnamefont {A.}~\bibnamefont {Vishwanath}}, \bibinfo {author} {\bibfnamefont {H.~C.}\ \bibnamefont {Po}}, \ and\ \bibinfo {author} {\bibfnamefont {X.}~\bibnamefont {Wan}},\ }\href {\doibase 10.1103/PhysRevB.100.195108} {\bibfield  {journal} {\bibinfo  {journal} {Phys. Rev. B}\ }\textbf {\bibinfo {volume} {100}},\ \bibinfo {pages} {195108} (\bibinfo {year} {2019})}\BibitemShut {NoStop}%
\bibitem [{\citenamefont {Marrazzo}\ \emph {et~al.}(2019)\citenamefont {Marrazzo}, \citenamefont {Gibertini}, \citenamefont {Campi}, \citenamefont {Mounet},\ and\ \citenamefont {Marzari}}]{marrazzo2019relative}%
  \BibitemOpen
  \bibfield  {author} {\bibinfo {author} {\bibfnamefont {A.}~\bibnamefont {Marrazzo}}, \bibinfo {author} {\bibfnamefont {M.}~\bibnamefont {Gibertini}}, \bibinfo {author} {\bibfnamefont {D.}~\bibnamefont {Campi}}, \bibinfo {author} {\bibfnamefont {N.}~\bibnamefont {Mounet}}, \ and\ \bibinfo {author} {\bibfnamefont {N.}~\bibnamefont {Marzari}},\ }\href {\doibase https://doi.org/10.1021/acs.nanolett.9b02689} {\bibfield  {journal} {\bibinfo  {journal} {Nano letters}\ }\textbf {\bibinfo {volume} {19}},\ \bibinfo {pages} {8431} (\bibinfo {year} {2019})}\BibitemShut {NoStop}%
\bibitem [{\citenamefont {Petralanda}\ \emph {et~al.}(2024)\citenamefont {Petralanda}, \citenamefont {Jiang}, \citenamefont {Bernevig}, \citenamefont {Regnault},\ and\ \citenamefont {Elcoro}}]{petralanda2024two}%
  \BibitemOpen
  \bibfield  {author} {\bibinfo {author} {\bibfnamefont {U.}~\bibnamefont {Petralanda}}, \bibinfo {author} {\bibfnamefont {Y.}~\bibnamefont {Jiang}}, \bibinfo {author} {\bibfnamefont {B.~A.}\ \bibnamefont {Bernevig}}, \bibinfo {author} {\bibfnamefont {N.}~\bibnamefont {Regnault}}, \ and\ \bibinfo {author} {\bibfnamefont {L.}~\bibnamefont {Elcoro}},\ }\href@noop {} {\bibfield  {journal} {\bibinfo  {journal} {arXiv:2411.08950}\ } (\bibinfo {year} {2024})}\BibitemShut {NoStop}%
\bibitem [{\citenamefont {Batatia}\ \emph {et~al.}(2022)\citenamefont {Batatia}, \citenamefont {Kovacs}, \citenamefont {Simm}, \citenamefont {Ortner},\ and\ \citenamefont {Cs{\'a}nyi}}]{batatia2022mace}%
  \BibitemOpen
  \bibfield  {author} {\bibinfo {author} {\bibfnamefont {I.}~\bibnamefont {Batatia}}, \bibinfo {author} {\bibfnamefont {D.~P.}\ \bibnamefont {Kovacs}}, \bibinfo {author} {\bibfnamefont {G.}~\bibnamefont {Simm}}, \bibinfo {author} {\bibfnamefont {C.}~\bibnamefont {Ortner}}, \ and\ \bibinfo {author} {\bibfnamefont {G.}~\bibnamefont {Cs{\'a}nyi}},\ }\href@noop {} {\bibfield  {journal} {\bibinfo  {journal} {Advances in neural information processing systems}\ }\textbf {\bibinfo {volume} {35}},\ \bibinfo {pages} {11423} (\bibinfo {year} {2022})}\BibitemShut {NoStop}%
\bibitem [{\citenamefont {Deng}\ \emph {et~al.}(2023)\citenamefont {Deng}, \citenamefont {Zhong}, \citenamefont {Jun}, \citenamefont {Riebesell}, \citenamefont {Han}, \citenamefont {Bartel},\ and\ \citenamefont {Ceder}}]{deng2023chgnet}%
  \BibitemOpen
  \bibfield  {author} {\bibinfo {author} {\bibfnamefont {B.}~\bibnamefont {Deng}}, \bibinfo {author} {\bibfnamefont {P.}~\bibnamefont {Zhong}}, \bibinfo {author} {\bibfnamefont {K.}~\bibnamefont {Jun}}, \bibinfo {author} {\bibfnamefont {J.}~\bibnamefont {Riebesell}}, \bibinfo {author} {\bibfnamefont {K.}~\bibnamefont {Han}}, \bibinfo {author} {\bibfnamefont {C.~J.}\ \bibnamefont {Bartel}}, \ and\ \bibinfo {author} {\bibfnamefont {G.}~\bibnamefont {Ceder}},\ }\href {\doibase https://doi.org/10.1038/s42256-023-00716-3} {\bibfield  {journal} {\bibinfo  {journal} {Nature Machine Intelligence}\ }\textbf {\bibinfo {volume} {5}},\ \bibinfo {pages} {1031} (\bibinfo {year} {2023})}\BibitemShut {NoStop}%
\bibitem [{\citenamefont {Chen}\ and\ \citenamefont {Ong}(2022)}]{chen2022universal}%
  \BibitemOpen
  \bibfield  {author} {\bibinfo {author} {\bibfnamefont {C.}~\bibnamefont {Chen}}\ and\ \bibinfo {author} {\bibfnamefont {S.~P.}\ \bibnamefont {Ong}},\ }\href {\doibase https://doi.org/10.1038/s43588-022-00349-3} {\bibfield  {journal} {\bibinfo  {journal} {Nature Computational Science}\ }\textbf {\bibinfo {volume} {2}},\ \bibinfo {pages} {718} (\bibinfo {year} {2022})}\BibitemShut {NoStop}%
\bibitem [{\citenamefont {Tyner}(2024{\natexlab{a}})}]{TynerML}%
  \BibitemOpen
  \bibfield  {author} {\bibinfo {author} {\bibfnamefont {A.~C.}\ \bibnamefont {Tyner}},\ }\href {\doibase 10.1103/PhysRevResearch.6.023316} {\bibfield  {journal} {\bibinfo  {journal} {Phys. Rev. Res.}\ }\textbf {\bibinfo {volume} {6}},\ \bibinfo {pages} {023316} (\bibinfo {year} {2024}{\natexlab{a}})}\BibitemShut {NoStop}%
\bibitem [{\citenamefont {Hong}\ \emph {et~al.}(2025)\citenamefont {Hong}, \citenamefont {Chen}, \citenamefont {Jin}, \citenamefont {Zhu}, \citenamefont {Gao}, \citenamefont {Zhao}, \citenamefont {Zhang}, \citenamefont {Ren},\ and\ \citenamefont {Cao}}]{hong2025discovery}%
  \BibitemOpen
  \bibfield  {author} {\bibinfo {author} {\bibfnamefont {T.}~\bibnamefont {Hong}}, \bibinfo {author} {\bibfnamefont {T.}~\bibnamefont {Chen}}, \bibinfo {author} {\bibfnamefont {D.}~\bibnamefont {Jin}}, \bibinfo {author} {\bibfnamefont {Y.}~\bibnamefont {Zhu}}, \bibinfo {author} {\bibfnamefont {H.}~\bibnamefont {Gao}}, \bibinfo {author} {\bibfnamefont {K.}~\bibnamefont {Zhao}}, \bibinfo {author} {\bibfnamefont {T.}~\bibnamefont {Zhang}}, \bibinfo {author} {\bibfnamefont {W.}~\bibnamefont {Ren}}, \ and\ \bibinfo {author} {\bibfnamefont {G.}~\bibnamefont {Cao}},\ }\href {\doibase https://doi.org/10.1038/s41535-025-00731-0} {\bibfield  {journal} {\bibinfo  {journal} {npj Quantum Materials}\ }\textbf {\bibinfo {volume} {10}},\ \bibinfo {pages} {12} (\bibinfo {year} {2025})}\BibitemShut {NoStop}%
\bibitem [{\citenamefont {Xie}\ and\ \citenamefont {Grossman}(2018)}]{cgcnn}%
  \BibitemOpen
  \bibfield  {author} {\bibinfo {author} {\bibfnamefont {T.}~\bibnamefont {Xie}}\ and\ \bibinfo {author} {\bibfnamefont {J.~C.}\ \bibnamefont {Grossman}},\ }\href {\doibase 10.1103/PhysRevLett.120.145301} {\bibfield  {journal} {\bibinfo  {journal} {Phys. Rev. Lett.}\ }\textbf {\bibinfo {volume} {120}},\ \bibinfo {pages} {145301} (\bibinfo {year} {2018})}\BibitemShut {NoStop}%
\bibitem [{\citenamefont {Park}\ and\ \citenamefont {Wolverton}(2020)}]{cgcnn_wolverton}%
  \BibitemOpen
  \bibfield  {author} {\bibinfo {author} {\bibfnamefont {C.~W.}\ \bibnamefont {Park}}\ and\ \bibinfo {author} {\bibfnamefont {C.}~\bibnamefont {Wolverton}},\ }\href {\doibase 10.1103/PhysRevMaterials.4.063801} {\bibfield  {journal} {\bibinfo  {journal} {Phys. Rev. Mater.}\ }\textbf {\bibinfo {volume} {4}},\ \bibinfo {pages} {063801} (\bibinfo {year} {2020})}\BibitemShut {NoStop}%
\bibitem [{\citenamefont {Thouless}\ \emph {et~al.}(1982)\citenamefont {Thouless}, \citenamefont {Kohmoto}, \citenamefont {Nightingale},\ and\ \citenamefont {den Nijs}}]{TKNN1982}%
  \BibitemOpen
  \bibfield  {author} {\bibinfo {author} {\bibfnamefont {D.~J.}\ \bibnamefont {Thouless}}, \bibinfo {author} {\bibfnamefont {M.}~\bibnamefont {Kohmoto}}, \bibinfo {author} {\bibfnamefont {M.~P.}\ \bibnamefont {Nightingale}}, \ and\ \bibinfo {author} {\bibfnamefont {M.}~\bibnamefont {den Nijs}},\ }\href {\doibase 10.1103/PhysRevLett.49.405} {\bibfield  {journal} {\bibinfo  {journal} {Phys. Rev. Lett.}\ }\textbf {\bibinfo {volume} {49}},\ \bibinfo {pages} {405} (\bibinfo {year} {1982})}\BibitemShut {NoStop}%
\bibitem [{\citenamefont {Niu}\ \emph {et~al.}(1985)\citenamefont {Niu}, \citenamefont {Thouless},\ and\ \citenamefont {Wu}}]{Niu1985}%
  \BibitemOpen
  \bibfield  {author} {\bibinfo {author} {\bibfnamefont {Q.}~\bibnamefont {Niu}}, \bibinfo {author} {\bibfnamefont {D.~J.}\ \bibnamefont {Thouless}}, \ and\ \bibinfo {author} {\bibfnamefont {Y.-S.}\ \bibnamefont {Wu}},\ }\href {\doibase 10.1103/PhysRevB.31.3372} {\bibfield  {journal} {\bibinfo  {journal} {Phys. Rev. B}\ }\textbf {\bibinfo {volume} {31}},\ \bibinfo {pages} {3372} (\bibinfo {year} {1985})}\BibitemShut {NoStop}%
\bibitem [{\citenamefont {Fu}\ and\ \citenamefont {Kane}(2007)}]{FuKane}%
  \BibitemOpen
  \bibfield  {author} {\bibinfo {author} {\bibfnamefont {L.}~\bibnamefont {Fu}}\ and\ \bibinfo {author} {\bibfnamefont {C.~L.}\ \bibnamefont {Kane}},\ }\href {\doibase 10.1103/PhysRevB.76.045302} {\bibfield  {journal} {\bibinfo  {journal} {Phys. Rev. B}\ }\textbf {\bibinfo {volume} {76}},\ \bibinfo {pages} {045302} (\bibinfo {year} {2007})}\BibitemShut {NoStop}%
\bibitem [{\citenamefont {Kane}\ and\ \citenamefont {Mele}(2005)}]{Kane2005}%
  \BibitemOpen
  \bibfield  {author} {\bibinfo {author} {\bibfnamefont {C.~L.}\ \bibnamefont {Kane}}\ and\ \bibinfo {author} {\bibfnamefont {E.~J.}\ \bibnamefont {Mele}},\ }\href {\doibase 10.1103/PhysRevLett.95.146802} {\bibfield  {journal} {\bibinfo  {journal} {Phys. Rev. Lett.}\ }\textbf {\bibinfo {volume} {95}},\ \bibinfo {pages} {146802} (\bibinfo {year} {2005})}\BibitemShut {NoStop}%
\bibitem [{\citenamefont {Sheng}\ \emph {et~al.}(2006)\citenamefont {Sheng}, \citenamefont {Weng}, \citenamefont {Sheng},\ and\ \citenamefont {Haldane}}]{Sheng2006}%
  \BibitemOpen
  \bibfield  {author} {\bibinfo {author} {\bibfnamefont {D.~N.}\ \bibnamefont {Sheng}}, \bibinfo {author} {\bibfnamefont {Z.~Y.}\ \bibnamefont {Weng}}, \bibinfo {author} {\bibfnamefont {L.}~\bibnamefont {Sheng}}, \ and\ \bibinfo {author} {\bibfnamefont {F.~D.~M.}\ \bibnamefont {Haldane}},\ }\href {\doibase 10.1103/PhysRevLett.97.036808} {\bibfield  {journal} {\bibinfo  {journal} {Phys. Rev. Lett.}\ }\textbf {\bibinfo {volume} {97}},\ \bibinfo {pages} {036808} (\bibinfo {year} {2006})}\BibitemShut {NoStop}%
\bibitem [{\citenamefont {Bernevig}\ \emph {et~al.}(2006)\citenamefont {Bernevig}, \citenamefont {Hughes},\ and\ \citenamefont {Zhang}}]{bernevig2006quantum}%
  \BibitemOpen
  \bibfield  {author} {\bibinfo {author} {\bibfnamefont {B.~A.}\ \bibnamefont {Bernevig}}, \bibinfo {author} {\bibfnamefont {T.~L.}\ \bibnamefont {Hughes}}, \ and\ \bibinfo {author} {\bibfnamefont {S.-C.}\ \bibnamefont {Zhang}},\ }\href {\doibase 10.1126/science.1133734} {\bibfield  {journal} {\bibinfo  {journal} {Science}\ }\textbf {\bibinfo {volume} {314}},\ \bibinfo {pages} {1757} (\bibinfo {year} {2006})}\BibitemShut {NoStop}%
\bibitem [{\citenamefont {Prodan}(2009)}]{Prodan2009}%
  \BibitemOpen
  \bibfield  {author} {\bibinfo {author} {\bibfnamefont {E.}~\bibnamefont {Prodan}},\ }\href {\doibase 10.1103/PhysRevB.80.125327} {\bibfield  {journal} {\bibinfo  {journal} {Phys. Rev. B}\ }\textbf {\bibinfo {volume} {80}},\ \bibinfo {pages} {125327} (\bibinfo {year} {2009})}\BibitemShut {NoStop}%
\bibitem [{\citenamefont {Benalcazar}\ \emph {et~al.}(2017)\citenamefont {Benalcazar}, \citenamefont {Bernevig},\ and\ \citenamefont {Hughes}}]{Benalcazar61}%
  \BibitemOpen
  \bibfield  {author} {\bibinfo {author} {\bibfnamefont {W.~A.}\ \bibnamefont {Benalcazar}}, \bibinfo {author} {\bibfnamefont {B.~A.}\ \bibnamefont {Bernevig}}, \ and\ \bibinfo {author} {\bibfnamefont {T.~L.}\ \bibnamefont {Hughes}},\ }\href {\doibase 10.1126/science.aah6442} {\bibfield  {journal} {\bibinfo  {journal} {Science}\ }\textbf {\bibinfo {volume} {357}},\ \bibinfo {pages} {61} (\bibinfo {year} {2017})}\BibitemShut {NoStop}%
\bibitem [{\citenamefont {Benalcazar}\ \emph {et~al.}(2019)\citenamefont {Benalcazar}, \citenamefont {Li},\ and\ \citenamefont {Hughes}}]{BenalacazarCn}%
  \BibitemOpen
  \bibfield  {author} {\bibinfo {author} {\bibfnamefont {W.~A.}\ \bibnamefont {Benalcazar}}, \bibinfo {author} {\bibfnamefont {T.}~\bibnamefont {Li}}, \ and\ \bibinfo {author} {\bibfnamefont {T.~L.}\ \bibnamefont {Hughes}},\ }\href {\doibase 10.1103/PhysRevB.99.245151} {\bibfield  {journal} {\bibinfo  {journal} {Phys. Rev. B}\ }\textbf {\bibinfo {volume} {99}},\ \bibinfo {pages} {245151} (\bibinfo {year} {2019})}\BibitemShut {NoStop}%
\bibitem [{\citenamefont {Schindler}\ \emph {et~al.}(2018{\natexlab{a}})\citenamefont {Schindler}, \citenamefont {Wang}, \citenamefont {Vergniory}, \citenamefont {Cook}, \citenamefont {Murani}, \citenamefont {Sengupta}, \citenamefont {Kasumov}, \citenamefont {Deblock}, \citenamefont {Jeon}, \citenamefont {Drozdov} \emph {et~al.}}]{schindler2018higher}%
  \BibitemOpen
  \bibfield  {author} {\bibinfo {author} {\bibfnamefont {F.}~\bibnamefont {Schindler}}, \bibinfo {author} {\bibfnamefont {Z.}~\bibnamefont {Wang}}, \bibinfo {author} {\bibfnamefont {M.~G.}\ \bibnamefont {Vergniory}}, \bibinfo {author} {\bibfnamefont {A.~M.}\ \bibnamefont {Cook}}, \bibinfo {author} {\bibfnamefont {A.}~\bibnamefont {Murani}}, \bibinfo {author} {\bibfnamefont {S.}~\bibnamefont {Sengupta}}, \bibinfo {author} {\bibfnamefont {A.~Y.}\ \bibnamefont {Kasumov}}, \bibinfo {author} {\bibfnamefont {R.}~\bibnamefont {Deblock}}, \bibinfo {author} {\bibfnamefont {S.}~\bibnamefont {Jeon}}, \bibinfo {author} {\bibfnamefont {I.}~\bibnamefont {Drozdov}},  \emph {et~al.},\ }\href {\doibase 10.1038/s41567-020-0902-0} {\bibfield  {journal} {\bibinfo  {journal} {Nat. Phys.}\ }\textbf {\bibinfo {volume} {14}},\ \bibinfo {pages} {918} (\bibinfo {year} {2018}{\natexlab{a}})}\BibitemShut {NoStop}%
\bibitem [{\citenamefont {Schindler}\ \emph {et~al.}(2018{\natexlab{b}})\citenamefont {Schindler}, \citenamefont {Cook}, \citenamefont {Vergniory}, \citenamefont {Wang}, \citenamefont {Parkin}, \citenamefont {Bernevig},\ and\ \citenamefont {Neupert}}]{Schindlereaat0346}%
  \BibitemOpen
  \bibfield  {author} {\bibinfo {author} {\bibfnamefont {F.}~\bibnamefont {Schindler}}, \bibinfo {author} {\bibfnamefont {A.~M.}\ \bibnamefont {Cook}}, \bibinfo {author} {\bibfnamefont {M.~G.}\ \bibnamefont {Vergniory}}, \bibinfo {author} {\bibfnamefont {Z.}~\bibnamefont {Wang}}, \bibinfo {author} {\bibfnamefont {S.~S.~P.}\ \bibnamefont {Parkin}}, \bibinfo {author} {\bibfnamefont {B.~A.}\ \bibnamefont {Bernevig}}, \ and\ \bibinfo {author} {\bibfnamefont {T.}~\bibnamefont {Neupert}},\ }\href {\doibase 10.1126/sciadv.aat0346} {\bibfield  {journal} {\bibinfo  {journal} {Sci. Adv.}\ }\textbf {\bibinfo {volume} {4}} (\bibinfo {year} {2018}{\natexlab{b}}),\ 10.1126/sciadv.aat0346}\BibitemShut {NoStop}%
\bibitem [{\citenamefont {Song}\ \emph {et~al.}(2020)\citenamefont {Song}, \citenamefont {Elcoro},\ and\ \citenamefont {Bernevig}}]{song2020twisted}%
  \BibitemOpen
  \bibfield  {author} {\bibinfo {author} {\bibfnamefont {Z.-D.}\ \bibnamefont {Song}}, \bibinfo {author} {\bibfnamefont {L.}~\bibnamefont {Elcoro}}, \ and\ \bibinfo {author} {\bibfnamefont {B.~A.}\ \bibnamefont {Bernevig}},\ }\href {\doibase 10.1126/science.aaz7650} {\bibfield  {journal} {\bibinfo  {journal} {Science}\ }\textbf {\bibinfo {volume} {367}},\ \bibinfo {pages} {794} (\bibinfo {year} {2020})}\BibitemShut {NoStop}%
\bibitem [{\citenamefont {Po}\ \emph {et~al.}(2018)\citenamefont {Po}, \citenamefont {Watanabe},\ and\ \citenamefont {Vishwanath}}]{Fragile}%
  \BibitemOpen
  \bibfield  {author} {\bibinfo {author} {\bibfnamefont {H.~C.}\ \bibnamefont {Po}}, \bibinfo {author} {\bibfnamefont {H.}~\bibnamefont {Watanabe}}, \ and\ \bibinfo {author} {\bibfnamefont {A.}~\bibnamefont {Vishwanath}},\ }\href {\doibase 10.1103/PhysRevLett.121.126402} {\bibfield  {journal} {\bibinfo  {journal} {Phys. Rev. Lett.}\ }\textbf {\bibinfo {volume} {121}},\ \bibinfo {pages} {126402} (\bibinfo {year} {2018})}\BibitemShut {NoStop}%
\bibitem [{\citenamefont {Tyner}\ \emph {et~al.}(2023)\citenamefont {Tyner}, \citenamefont {Sur}, \citenamefont {Puggioni}, \citenamefont {Rondinelli},\ and\ \citenamefont {Goswami}}]{tyner2020topology}%
  \BibitemOpen
  \bibfield  {author} {\bibinfo {author} {\bibfnamefont {A.~C.}\ \bibnamefont {Tyner}}, \bibinfo {author} {\bibfnamefont {S.}~\bibnamefont {Sur}}, \bibinfo {author} {\bibfnamefont {D.}~\bibnamefont {Puggioni}}, \bibinfo {author} {\bibfnamefont {J.~M.}\ \bibnamefont {Rondinelli}}, \ and\ \bibinfo {author} {\bibfnamefont {P.}~\bibnamefont {Goswami}},\ }\href {\doibase 10.1103/PhysRevResearch.5.L012019} {\bibfield  {journal} {\bibinfo  {journal} {Phys. Rev. Res.}\ }\textbf {\bibinfo {volume} {5}},\ \bibinfo {pages} {L012019} (\bibinfo {year} {2023})}\BibitemShut {NoStop}%
\bibitem [{\citenamefont {Qi}\ and\ \citenamefont {Zhang}(2008)}]{QiSpinCharge}%
  \BibitemOpen
  \bibfield  {author} {\bibinfo {author} {\bibfnamefont {X.-L.}\ \bibnamefont {Qi}}\ and\ \bibinfo {author} {\bibfnamefont {S.-C.}\ \bibnamefont {Zhang}},\ }\href {\doibase 10.1103/PhysRevLett.101.086802} {\bibfield  {journal} {\bibinfo  {journal} {Phys. Rev. Lett.}\ }\textbf {\bibinfo {volume} {101}},\ \bibinfo {pages} {086802} (\bibinfo {year} {2008})}\BibitemShut {NoStop}%
\bibitem [{\citenamefont {Ran}\ \emph {et~al.}(2008)\citenamefont {Ran}, \citenamefont {Vishwanath},\ and\ \citenamefont {Lee}}]{SpinChargeVishwanath}%
  \BibitemOpen
  \bibfield  {author} {\bibinfo {author} {\bibfnamefont {Y.}~\bibnamefont {Ran}}, \bibinfo {author} {\bibfnamefont {A.}~\bibnamefont {Vishwanath}}, \ and\ \bibinfo {author} {\bibfnamefont {D.-H.}\ \bibnamefont {Lee}},\ }\href {\doibase 10.1103/PhysRevLett.101.086801} {\bibfield  {journal} {\bibinfo  {journal} {Phys. Rev. Lett.}\ }\textbf {\bibinfo {volume} {101}},\ \bibinfo {pages} {086801} (\bibinfo {year} {2008})}\BibitemShut {NoStop}%
\bibitem [{\citenamefont {Wang}\ and\ \citenamefont {Zhang}(2010)}]{Wang_2010}%
  \BibitemOpen
  \bibfield  {author} {\bibinfo {author} {\bibfnamefont {Z.}~\bibnamefont {Wang}}\ and\ \bibinfo {author} {\bibfnamefont {P.}~\bibnamefont {Zhang}},\ }\href {\doibase 10.1088/1367-2630/12/4/043055} {\bibfield  {journal} {\bibinfo  {journal} {New Journal of Physics}\ }\textbf {\bibinfo {volume} {12}},\ \bibinfo {pages} {043055} (\bibinfo {year} {2010})}\BibitemShut {NoStop}%
\bibitem [{\citenamefont {Juri\ifmmode \check{c}\else \v{c}\fi{}i\ifmmode~\acute{c}\else \'{c}\fi{}}\ \emph {et~al.}(2012)\citenamefont {Juri\ifmmode \check{c}\else \v{c}\fi{}i\ifmmode~\acute{c}\else \'{c}\fi{}}, \citenamefont {Mesaros}, \citenamefont {Slager},\ and\ \citenamefont {Zaanen}}]{slager2012}%
  \BibitemOpen
  \bibfield  {author} {\bibinfo {author} {\bibfnamefont {V.}~\bibnamefont {Juri\ifmmode \check{c}\else \v{c}\fi{}i\ifmmode~\acute{c}\else \'{c}\fi{}}}, \bibinfo {author} {\bibfnamefont {A.}~\bibnamefont {Mesaros}}, \bibinfo {author} {\bibfnamefont {R.-J.}\ \bibnamefont {Slager}}, \ and\ \bibinfo {author} {\bibfnamefont {J.}~\bibnamefont {Zaanen}},\ }\href {\doibase 10.1103/PhysRevLett.108.106403} {\bibfield  {journal} {\bibinfo  {journal} {Phys. Rev. Lett.}\ }\textbf {\bibinfo {volume} {108}},\ \bibinfo {pages} {106403} (\bibinfo {year} {2012})}\BibitemShut {NoStop}%
\bibitem [{\citenamefont {Mesaros}\ \emph {et~al.}(2013)\citenamefont {Mesaros}, \citenamefont {Slager}, \citenamefont {Zaanen},\ and\ \citenamefont {Juri{\v{c}}i{\'c}}}]{MESAROS2013977}%
  \BibitemOpen
  \bibfield  {author} {\bibinfo {author} {\bibfnamefont {A.}~\bibnamefont {Mesaros}}, \bibinfo {author} {\bibfnamefont {R.-J.}\ \bibnamefont {Slager}}, \bibinfo {author} {\bibfnamefont {J.}~\bibnamefont {Zaanen}}, \ and\ \bibinfo {author} {\bibfnamefont {V.}~\bibnamefont {Juri{\v{c}}i{\'c}}},\ }\href {\doibase 10.1016/j.nuclphysb.2012.10.022} {\bibfield  {journal} {\bibinfo  {journal} {Nuc. Phys. B}\ }\textbf {\bibinfo {volume} {867}},\ \bibinfo {pages} {977} (\bibinfo {year} {2013})}\BibitemShut {NoStop}%
\bibitem [{\citenamefont {Tyner}\ and\ \citenamefont {Goswami}(2023)}]{tynerbismuthene}%
  \BibitemOpen
  \bibfield  {author} {\bibinfo {author} {\bibfnamefont {A.~C.}\ \bibnamefont {Tyner}}\ and\ \bibinfo {author} {\bibfnamefont {P.}~\bibnamefont {Goswami}},\ }\href {\doibase https://doi.org/10.1038/s41598-023-38491-1} {\bibfield  {journal} {\bibinfo  {journal} {Sci. Rep.}\ }\textbf {\bibinfo {volume} {13}},\ \bibinfo {pages} {11393} (\bibinfo {year} {2023})}\BibitemShut {NoStop}%
\bibitem [{\citenamefont {Tyner}\ and\ \citenamefont {Goswami}(2024)}]{TynerRealSpace}%
  \BibitemOpen
  \bibfield  {author} {\bibinfo {author} {\bibfnamefont {A.~C.}\ \bibnamefont {Tyner}}\ and\ \bibinfo {author} {\bibfnamefont {P.}~\bibnamefont {Goswami}},\ }\href {\doibase 10.1103/PhysRevMaterials.8.124203} {\bibfield  {journal} {\bibinfo  {journal} {Phys. Rev. Mater.}\ }\textbf {\bibinfo {volume} {8}},\ \bibinfo {pages} {124203} (\bibinfo {year} {2024})}\BibitemShut {NoStop}%
\bibitem [{\citenamefont {Pizzi}\ \emph {et~al.}(2020)\citenamefont {Pizzi}, \citenamefont {Vitale}, \citenamefont {Arita}, \citenamefont {Blugel}, \citenamefont {Freimuth}, \citenamefont {G{\'{e}}ranton}, \citenamefont {Gibertini}, \citenamefont {Gresch}, \citenamefont {Johnson}, \citenamefont {Koretsune}, \citenamefont {Iba{\~{n}}ez-Azpiroz}, \citenamefont {Lee}, \citenamefont {Lihm}, \citenamefont {Marchand}, \citenamefont {Marrazzo}, \citenamefont {Mokrousov}, \citenamefont {Mustafa}, \citenamefont {Nohara}, \citenamefont {Nomura}, \citenamefont {Paulatto}, \citenamefont {Ponc{\'{e}}}, \citenamefont {Ponweiser}, \citenamefont {Qiao}, \citenamefont {Thole}, \citenamefont {Tsirkin}, \citenamefont {Wierzbowska}, \citenamefont {Marzari}, \citenamefont {Vanderbilt}, \citenamefont {Souza}, \citenamefont {Mostofi},\ and\ \citenamefont {Yates}}]{Pizzi2020}%
  \BibitemOpen
  \bibfield  {author} {\bibinfo {author} {\bibfnamefont {G.}~\bibnamefont {Pizzi}}, \bibinfo {author} {\bibfnamefont {V.}~\bibnamefont {Vitale}}, \bibinfo {author} {\bibfnamefont {R.}~\bibnamefont {Arita}}, \bibinfo {author} {\bibfnamefont {S.}~\bibnamefont {Blugel}}, \bibinfo {author} {\bibfnamefont {F.}~\bibnamefont {Freimuth}}, \bibinfo {author} {\bibfnamefont {G.}~\bibnamefont {G{\'{e}}ranton}}, \bibinfo {author} {\bibfnamefont {M.}~\bibnamefont {Gibertini}}, \bibinfo {author} {\bibfnamefont {D.}~\bibnamefont {Gresch}}, \bibinfo {author} {\bibfnamefont {C.}~\bibnamefont {Johnson}}, \bibinfo {author} {\bibfnamefont {T.}~\bibnamefont {Koretsune}}, \bibinfo {author} {\bibfnamefont {J.}~\bibnamefont {Iba{\~{n}}ez-Azpiroz}}, \bibinfo {author} {\bibfnamefont {H.}~\bibnamefont {Lee}}, \bibinfo {author} {\bibfnamefont {J.-M.}\ \bibnamefont {Lihm}}, \bibinfo {author} {\bibfnamefont {D.}~\bibnamefont {Marchand}}, \bibinfo {author} {\bibfnamefont {A.}~\bibnamefont {Marrazzo}}, \bibinfo {author} {\bibfnamefont
  {Y.}~\bibnamefont {Mokrousov}}, \bibinfo {author} {\bibfnamefont {J.~I.}\ \bibnamefont {Mustafa}}, \bibinfo {author} {\bibfnamefont {Y.}~\bibnamefont {Nohara}}, \bibinfo {author} {\bibfnamefont {Y.}~\bibnamefont {Nomura}}, \bibinfo {author} {\bibfnamefont {L.}~\bibnamefont {Paulatto}}, \bibinfo {author} {\bibfnamefont {S.}~\bibnamefont {Ponc{\'{e}}}}, \bibinfo {author} {\bibfnamefont {T.}~\bibnamefont {Ponweiser}}, \bibinfo {author} {\bibfnamefont {J.}~\bibnamefont {Qiao}}, \bibinfo {author} {\bibfnamefont {F.}~\bibnamefont {Thole}}, \bibinfo {author} {\bibfnamefont {S.~S.}\ \bibnamefont {Tsirkin}}, \bibinfo {author} {\bibfnamefont {M.}~\bibnamefont {Wierzbowska}}, \bibinfo {author} {\bibfnamefont {N.}~\bibnamefont {Marzari}}, \bibinfo {author} {\bibfnamefont {D.}~\bibnamefont {Vanderbilt}}, \bibinfo {author} {\bibfnamefont {I.}~\bibnamefont {Souza}}, \bibinfo {author} {\bibfnamefont {A.~A.}\ \bibnamefont {Mostofi}}, \ and\ \bibinfo {author} {\bibfnamefont {J.~R.}\ \bibnamefont {Yates}},\ }\href {\doibase
  10.1088/1361-648x/ab51ff} {\bibfield  {journal} {\bibinfo  {journal} {J. Phys. Condens. Matter}\ }\textbf {\bibinfo {volume} {32}},\ \bibinfo {pages} {165902} (\bibinfo {year} {2020})}\BibitemShut {NoStop}%
\bibitem [{\citenamefont {Perdew}\ \emph {et~al.}(1997)\citenamefont {Perdew}, \citenamefont {Burke},\ and\ \citenamefont {Ernzerhof}}]{Perdew1996}%
  \BibitemOpen
  \bibfield  {author} {\bibinfo {author} {\bibfnamefont {J.~P.}\ \bibnamefont {Perdew}}, \bibinfo {author} {\bibfnamefont {K.}~\bibnamefont {Burke}}, \ and\ \bibinfo {author} {\bibfnamefont {M.}~\bibnamefont {Ernzerhof}},\ }\href {\doibase 10.1103/PhysRevLett.78.1396} {\bibfield  {journal} {\bibinfo  {journal} {Phys. Rev. Lett.}\ }\textbf {\bibinfo {volume} {78}},\ \bibinfo {pages} {1396} (\bibinfo {year} {1997})}\BibitemShut {NoStop}%
\bibitem [{\citenamefont {Van~Setten}\ \emph {et~al.}(2018)\citenamefont {Van~Setten}, \citenamefont {Giantomassi}, \citenamefont {Bousquet}, \citenamefont {Verstraete}, \citenamefont {Hamann}, \citenamefont {Gonze},\ and\ \citenamefont {Rignanese}}]{van2018pseudodojo}%
  \BibitemOpen
  \bibfield  {author} {\bibinfo {author} {\bibfnamefont {M.~J.}\ \bibnamefont {Van~Setten}}, \bibinfo {author} {\bibfnamefont {M.}~\bibnamefont {Giantomassi}}, \bibinfo {author} {\bibfnamefont {E.}~\bibnamefont {Bousquet}}, \bibinfo {author} {\bibfnamefont {M.~J.}\ \bibnamefont {Verstraete}}, \bibinfo {author} {\bibfnamefont {D.~R.}\ \bibnamefont {Hamann}}, \bibinfo {author} {\bibfnamefont {X.}~\bibnamefont {Gonze}}, \ and\ \bibinfo {author} {\bibfnamefont {G.-M.}\ \bibnamefont {Rignanese}},\ }\href {\doibase https://doi.org/10.1016/j.cpc.2018.01.012} {\bibfield  {journal} {\bibinfo  {journal} {Computer Physics Communications}\ }\textbf {\bibinfo {volume} {226}},\ \bibinfo {pages} {39} (\bibinfo {year} {2018})}\BibitemShut {NoStop}%
\bibitem [{\citenamefont {Tyner}(2024{\natexlab{b}})}]{tyner2023berryeasy}%
  \BibitemOpen
  \bibfield  {author} {\bibinfo {author} {\bibfnamefont {A.~C.}\ \bibnamefont {Tyner}},\ }\href {\doibase 10.1088/1361-648X/ad475f} {\bibfield  {journal} {\bibinfo  {journal} {J. Condens. Matter Phys.}\ }\textbf {\bibinfo {volume} {36}},\ \bibinfo {pages} {325902} (\bibinfo {year} {2024}{\natexlab{b}})}\BibitemShut {NoStop}%
\bibitem [{\citenamefont {Wu}\ \emph {et~al.}(2018)\citenamefont {Wu}, \citenamefont {Zhang}, \citenamefont {Song}, \citenamefont {Troyer},\ and\ \citenamefont {Soluyanov}}]{WU2017}%
  \BibitemOpen
  \bibfield  {author} {\bibinfo {author} {\bibfnamefont {Q.}~\bibnamefont {Wu}}, \bibinfo {author} {\bibfnamefont {S.}~\bibnamefont {Zhang}}, \bibinfo {author} {\bibfnamefont {H.-F.}\ \bibnamefont {Song}}, \bibinfo {author} {\bibfnamefont {M.}~\bibnamefont {Troyer}}, \ and\ \bibinfo {author} {\bibfnamefont {A.~A.}\ \bibnamefont {Soluyanov}},\ }\href {\doibase https://doi.org/10.1016/j.cpc.2017.09.033} {\bibfield  {journal} {\bibinfo  {journal} {Computer Physics Communications}\ }\textbf {\bibinfo {volume} {224}},\ \bibinfo {pages} {405 } (\bibinfo {year} {2018})}\BibitemShut {NoStop}%
\bibitem [{git(2025)}]{github}%
  \BibitemOpen
  \href {https://github.com/actyner/GAN_MLP} {\  (\bibinfo {year} {2025})}\BibitemShut {NoStop}%
\bibitem [{\citenamefont {Qian}\ \emph {et~al.}(2014)\citenamefont {Qian}, \citenamefont {Liu}, \citenamefont {Fu},\ and\ \citenamefont {Li}}]{qian2014quantum}%
  \BibitemOpen
  \bibfield  {author} {\bibinfo {author} {\bibfnamefont {X.}~\bibnamefont {Qian}}, \bibinfo {author} {\bibfnamefont {J.}~\bibnamefont {Liu}}, \bibinfo {author} {\bibfnamefont {L.}~\bibnamefont {Fu}}, \ and\ \bibinfo {author} {\bibfnamefont {J.}~\bibnamefont {Li}},\ }\href {\doibase 10.1126/science.1256815} {\bibfield  {journal} {\bibinfo  {journal} {Science}\ }\textbf {\bibinfo {volume} {346}},\ \bibinfo {pages} {1344} (\bibinfo {year} {2014})}\BibitemShut {NoStop}%
\bibitem [{\citenamefont {Yu}\ \emph {et~al.}(2011)\citenamefont {Yu}, \citenamefont {Qi}, \citenamefont {Bernevig}, \citenamefont {Fang},\ and\ \citenamefont {Dai}}]{yu2011equivalent}%
  \BibitemOpen
  \bibfield  {author} {\bibinfo {author} {\bibfnamefont {R.}~\bibnamefont {Yu}}, \bibinfo {author} {\bibfnamefont {X.~L.}\ \bibnamefont {Qi}}, \bibinfo {author} {\bibfnamefont {A.}~\bibnamefont {Bernevig}}, \bibinfo {author} {\bibfnamefont {Z.}~\bibnamefont {Fang}}, \ and\ \bibinfo {author} {\bibfnamefont {X.}~\bibnamefont {Dai}},\ }\href {\doibase 10.1103/PhysRevB.84.075119} {\bibfield  {journal} {\bibinfo  {journal} {Phys. Rev. B}\ }\textbf {\bibinfo {volume} {84}},\ \bibinfo {pages} {075119} (\bibinfo {year} {2011})}\BibitemShut {NoStop}%
\bibitem [{\citenamefont {Gresch}\ \emph {et~al.}(2017)\citenamefont {Gresch}, \citenamefont {Aut\`es}, \citenamefont {Yazyev}, \citenamefont {Troyer}, \citenamefont {Vanderbilt}, \citenamefont {Bernevig},\ and\ \citenamefont {Soluyanov}}]{Z2pack}%
  \BibitemOpen
  \bibfield  {author} {\bibinfo {author} {\bibfnamefont {D.}~\bibnamefont {Gresch}}, \bibinfo {author} {\bibfnamefont {G.}~\bibnamefont {Aut\`es}}, \bibinfo {author} {\bibfnamefont {O.~V.}\ \bibnamefont {Yazyev}}, \bibinfo {author} {\bibfnamefont {M.}~\bibnamefont {Troyer}}, \bibinfo {author} {\bibfnamefont {D.}~\bibnamefont {Vanderbilt}}, \bibinfo {author} {\bibfnamefont {B.~A.}\ \bibnamefont {Bernevig}}, \ and\ \bibinfo {author} {\bibfnamefont {A.~A.}\ \bibnamefont {Soluyanov}},\ }\href {\doibase 10.1103/PhysRevB.95.075146} {\bibfield  {journal} {\bibinfo  {journal} {Phys. Rev. B}\ }\textbf {\bibinfo {volume} {95}},\ \bibinfo {pages} {075146} (\bibinfo {year} {2017})}\BibitemShut {NoStop}%
\bibitem [{\citenamefont {Taherinejad}\ \emph {et~al.}(2014)\citenamefont {Taherinejad}, \citenamefont {Garrity},\ and\ \citenamefont {Vanderbilt}}]{Taherinejad2014}%
  \BibitemOpen
  \bibfield  {author} {\bibinfo {author} {\bibfnamefont {M.}~\bibnamefont {Taherinejad}}, \bibinfo {author} {\bibfnamefont {K.~F.}\ \bibnamefont {Garrity}}, \ and\ \bibinfo {author} {\bibfnamefont {D.}~\bibnamefont {Vanderbilt}},\ }\href {\doibase 10.1103/PhysRevB.89.115102} {\bibfield  {journal} {\bibinfo  {journal} {Phys. Rev. B}\ }\textbf {\bibinfo {volume} {89}},\ \bibinfo {pages} {1} (\bibinfo {year} {2014})},\ \Eprint {http://arxiv.org/abs/1312.6940} {1312.6940} \BibitemShut {NoStop}%
\bibitem [{\citenamefont {Zhao}\ \emph {et~al.}(2016)\citenamefont {Zhao}, \citenamefont {Zhang}, \citenamefont {Ji}, \citenamefont {Zhang}, \citenamefont {Li}, \citenamefont {Yan}, \citenamefont {Zhang}, \citenamefont {Li},\ and\ \citenamefont {Wang}}]{zhao2016unexpected}%
  \BibitemOpen
  \bibfield  {author} {\bibinfo {author} {\bibfnamefont {H.}~\bibnamefont {Zhao}}, \bibinfo {author} {\bibfnamefont {C.-w.}\ \bibnamefont {Zhang}}, \bibinfo {author} {\bibfnamefont {W.-x.}\ \bibnamefont {Ji}}, \bibinfo {author} {\bibfnamefont {R.-w.}\ \bibnamefont {Zhang}}, \bibinfo {author} {\bibfnamefont {S.-s.}\ \bibnamefont {Li}}, \bibinfo {author} {\bibfnamefont {S.-s.}\ \bibnamefont {Yan}}, \bibinfo {author} {\bibfnamefont {B.-m.}\ \bibnamefont {Zhang}}, \bibinfo {author} {\bibfnamefont {P.}~\bibnamefont {Li}}, \ and\ \bibinfo {author} {\bibfnamefont {P.-j.}\ \bibnamefont {Wang}},\ }\href {\doibase https://doi.org/10.1038/srep20152} {\bibfield  {journal} {\bibinfo  {journal} {Scientific reports}\ }\textbf {\bibinfo {volume} {6}},\ \bibinfo {pages} {20152} (\bibinfo {year} {2016})}\BibitemShut {NoStop}%
\bibitem [{\citenamefont {Costa}\ \emph {et~al.}(2021)\citenamefont {Costa}, \citenamefont {Schleder}, \citenamefont {Mera~Acosta}, \citenamefont {Padilha}, \citenamefont {Cerasoli}, \citenamefont {Buongiorno~Nardelli},\ and\ \citenamefont {Fazzio}}]{costa2021discovery}%
  \BibitemOpen
  \bibfield  {author} {\bibinfo {author} {\bibfnamefont {M.}~\bibnamefont {Costa}}, \bibinfo {author} {\bibfnamefont {G.~R.}\ \bibnamefont {Schleder}}, \bibinfo {author} {\bibfnamefont {C.}~\bibnamefont {Mera~Acosta}}, \bibinfo {author} {\bibfnamefont {A.~C.}\ \bibnamefont {Padilha}}, \bibinfo {author} {\bibfnamefont {F.}~\bibnamefont {Cerasoli}}, \bibinfo {author} {\bibfnamefont {M.}~\bibnamefont {Buongiorno~Nardelli}}, \ and\ \bibinfo {author} {\bibfnamefont {A.}~\bibnamefont {Fazzio}},\ }\href {\doibase https://doi.org/10.1038/s41524-021-00518-4} {\bibfield  {journal} {\bibinfo  {journal} {npj Comput Mater}\ }\textbf {\bibinfo {volume} {7}},\ \bibinfo {pages} {49} (\bibinfo {year} {2021})}\BibitemShut {NoStop}%
\bibitem [{\citenamefont {Lin}\ \emph {et~al.}(2024)\citenamefont {Lin}, \citenamefont {Palumbo}, \citenamefont {Guo}, \citenamefont {Hwang}, \citenamefont {Blackburn}, \citenamefont {Shoemaker}, \citenamefont {Mahmood}, \citenamefont {Wang}, \citenamefont {Fiete}, \citenamefont {Wieder} \emph {et~al.}}]{Lin2022Spin}%
  \BibitemOpen
  \bibfield  {author} {\bibinfo {author} {\bibfnamefont {K.-S.}\ \bibnamefont {Lin}}, \bibinfo {author} {\bibfnamefont {G.}~\bibnamefont {Palumbo}}, \bibinfo {author} {\bibfnamefont {Z.}~\bibnamefont {Guo}}, \bibinfo {author} {\bibfnamefont {Y.}~\bibnamefont {Hwang}}, \bibinfo {author} {\bibfnamefont {J.}~\bibnamefont {Blackburn}}, \bibinfo {author} {\bibfnamefont {D.~P.}\ \bibnamefont {Shoemaker}}, \bibinfo {author} {\bibfnamefont {F.}~\bibnamefont {Mahmood}}, \bibinfo {author} {\bibfnamefont {Z.}~\bibnamefont {Wang}}, \bibinfo {author} {\bibfnamefont {G.~A.}\ \bibnamefont {Fiete}}, \bibinfo {author} {\bibfnamefont {B.~J.}\ \bibnamefont {Wieder}},  \emph {et~al.},\ }\href {\doibase https://doi.org/10.1038/s41467-024-44762-w} {\bibfield  {journal} {\bibinfo  {journal} {Nat. Commun.}\ }\textbf {\bibinfo {volume} {15}},\ \bibinfo {pages} {550} (\bibinfo {year} {2024})}\BibitemShut {NoStop}%
\bibitem [{\citenamefont {Kang}\ \emph {et~al.}(2024)\citenamefont {Kang}, \citenamefont {Qiu}, \citenamefont {Watanabe}, \citenamefont {Taniguchi}, \citenamefont {Shan},\ and\ \citenamefont {Mak}}]{kang2024double}%
  \BibitemOpen
  \bibfield  {author} {\bibinfo {author} {\bibfnamefont {K.}~\bibnamefont {Kang}}, \bibinfo {author} {\bibfnamefont {Y.}~\bibnamefont {Qiu}}, \bibinfo {author} {\bibfnamefont {K.}~\bibnamefont {Watanabe}}, \bibinfo {author} {\bibfnamefont {T.}~\bibnamefont {Taniguchi}}, \bibinfo {author} {\bibfnamefont {J.}~\bibnamefont {Shan}}, \ and\ \bibinfo {author} {\bibfnamefont {K.~F.}\ \bibnamefont {Mak}},\ }\href {\doibase https://doi.org/10.1021/acs.nanolett.4c05308} {\bibfield  {journal} {\bibinfo  {journal} {Nano Lett.}\ }\textbf {\bibinfo {volume} {24}},\ \bibinfo {pages} {14901} (\bibinfo {year} {2024})}\BibitemShut {NoStop}%
\bibitem [{\citenamefont {Bai}\ \emph {et~al.}(2022)\citenamefont {Bai}, \citenamefont {Cai}, \citenamefont {Mao}, \citenamefont {Li}, \citenamefont {Dai}, \citenamefont {Huang},\ and\ \citenamefont {Niu}}]{Antimonene}%
  \BibitemOpen
  \bibfield  {author} {\bibinfo {author} {\bibfnamefont {Y.}~\bibnamefont {Bai}}, \bibinfo {author} {\bibfnamefont {L.}~\bibnamefont {Cai}}, \bibinfo {author} {\bibfnamefont {N.}~\bibnamefont {Mao}}, \bibinfo {author} {\bibfnamefont {R.}~\bibnamefont {Li}}, \bibinfo {author} {\bibfnamefont {Y.}~\bibnamefont {Dai}}, \bibinfo {author} {\bibfnamefont {B.}~\bibnamefont {Huang}}, \ and\ \bibinfo {author} {\bibfnamefont {C.}~\bibnamefont {Niu}},\ }\href {\doibase 10.1103/PhysRevB.105.195142} {\bibfield  {journal} {\bibinfo  {journal} {Phys. Rev. B}\ }\textbf {\bibinfo {volume} {105}},\ \bibinfo {pages} {195142} (\bibinfo {year} {2022})}\BibitemShut {NoStop}%
\end{thebibliography}%


\begin{thebibliography}{4}%
\makeatletter
\providecommand \@ifxundefined [1]{%
 \@ifx{#1\undefined}
}%
\providecommand \@ifnum [1]{%
 \ifnum #1\expandafter \@firstoftwo
 \else \expandafter \@secondoftwo
 \fi
}%
\providecommand \@ifx [1]{%
 \ifx #1\expandafter \@firstoftwo
 \else \expandafter \@secondoftwo
 \fi
}%
\providecommand \natexlab [1]{#1}%
\providecommand \enquote  [1]{``#1''}%
\providecommand \bibnamefont  [1]{#1}%
\providecommand \bibfnamefont [1]{#1}%
\providecommand \citenamefont [1]{#1}%
\providecommand \href@noop [0]{\@secondoftwo}%
\providecommand \href [0]{\begingroup \@sanitize@url \@href}%
\providecommand \@href[1]{\@@startlink{#1}\@@href}%
\providecommand \@@href[1]{\endgroup#1\@@endlink}%
\providecommand \@sanitize@url [0]{\catcode `\\12\catcode `\$12\catcode `\&12\catcode `\#12\catcode `\^12\catcode `\_12\catcode `\%12\relax}%
\providecommand \@@startlink[1]{}%
\providecommand \@@endlink[0]{}%
\providecommand \url  [0]{\begingroup\@sanitize@url \@url }%
\providecommand \@url [1]{\endgroup\@href {#1}{\urlprefix }}%
\providecommand \urlprefix  [0]{URL }%
\providecommand \Eprint [0]{\href }%
\providecommand \doibase [0]{http://dx.doi.org/}%
\providecommand \selectlanguage [0]{\@gobble}%
\providecommand \bibinfo  [0]{\@secondoftwo}%
\providecommand \bibfield  [0]{\@secondoftwo}%
\providecommand \translation [1]{[#1]}%
\providecommand \BibitemOpen [0]{}%
\providecommand \bibitemStop [0]{}%
\providecommand \bibitemNoStop [0]{.\EOS\space}%
\providecommand \EOS [0]{\spacefactor3000\relax}%
\providecommand \BibitemShut  [1]{\csname bibitem#1\endcsname}%
\let\auto@bib@innerbib\@empty
\bibitem [{\citenamefont {Xie}\ and\ \citenamefont {Grossman}(2018)}]{cgcnn}%
  \BibitemOpen
  \bibfield  {author} {\bibinfo {author} {\bibfnamefont {T.}~\bibnamefont {Xie}}\ and\ \bibinfo {author} {\bibfnamefont {J.~C.}\ \bibnamefont {Grossman}},\ }\href {\doibase 10.1103/PhysRevLett.120.145301} {\bibfield  {journal} {\bibinfo  {journal} {Phys. Rev. Lett.}\ }\textbf {\bibinfo {volume} {120}},\ \bibinfo {pages} {145301} (\bibinfo {year} {2018})}\BibitemShut {NoStop}%
\bibitem [{\citenamefont {Deng}\ \emph {et~al.}(2023)\citenamefont {Deng}, \citenamefont {Zhong}, \citenamefont {Jun}, \citenamefont {Riebesell}, \citenamefont {Han}, \citenamefont {Bartel},\ and\ \citenamefont {Ceder}}]{deng2023chgnet}%
  \BibitemOpen
  \bibfield  {author} {\bibinfo {author} {\bibfnamefont {B.}~\bibnamefont {Deng}}, \bibinfo {author} {\bibfnamefont {P.}~\bibnamefont {Zhong}}, \bibinfo {author} {\bibfnamefont {K.}~\bibnamefont {Jun}}, \bibinfo {author} {\bibfnamefont {J.}~\bibnamefont {Riebesell}}, \bibinfo {author} {\bibfnamefont {K.}~\bibnamefont {Han}}, \bibinfo {author} {\bibfnamefont {C.~J.}\ \bibnamefont {Bartel}}, \ and\ \bibinfo {author} {\bibfnamefont {G.}~\bibnamefont {Ceder}},\ }\href {\doibase https://doi.org/10.1038/s42256-023-00716-3} {\bibfield  {journal} {\bibinfo  {journal} {Nature Machine Intelligence}\ }\textbf {\bibinfo {volume} {5}},\ \bibinfo {pages} {1031} (\bibinfo {year} {2023})}\BibitemShut {NoStop}%
\bibitem [{git(2025)}]{github}%
  \BibitemOpen
  \href {https://github.com/actyner/GAN_MLP} {\  (\bibinfo {year} {2025})}\BibitemShut {NoStop}%
\bibitem [{\citenamefont {Prodan}(2009)}]{Prodan2009}%
  \BibitemOpen
  \bibfield  {author} {\bibinfo {author} {\bibfnamefont {E.}~\bibnamefont {Prodan}},\ }\href {\doibase 10.1103/PhysRevB.80.125327} {\bibfield  {journal} {\bibinfo  {journal} {Phys. Rev. B}\ }\textbf {\bibinfo {volume} {80}},\ \bibinfo {pages} {125327} (\bibinfo {year} {2009})}\BibitemShut {NoStop}%
\end{thebibliography}%

\end{document}


\renewcommand{\theequation}{S\arabic{equation}}
\renewcommand{\thetable}{S\arabic{table}}
\makeatletter 
\renewcommand{\thefigure}{S\@arabic\c@figure}
\makeatother

\title{Supplementary Material: Fine tuning generative adversarial networks with universal force fields: application to two-dimensional topological insulators}
\author{Alexander C. Tyner}

\affiliation{NORDITA, KTH Royal Institute of Technology and Stockholm University 106 91 Stockholm, Sweden}
\affiliation{Department of Physics, University of Connecticut, Storrs, Connecticut 06269, USA}

\maketitle
\onecolumngrid

\section{Network Architecture}
\begin{figure*}
    \centering
    \includegraphics[width=16cm]{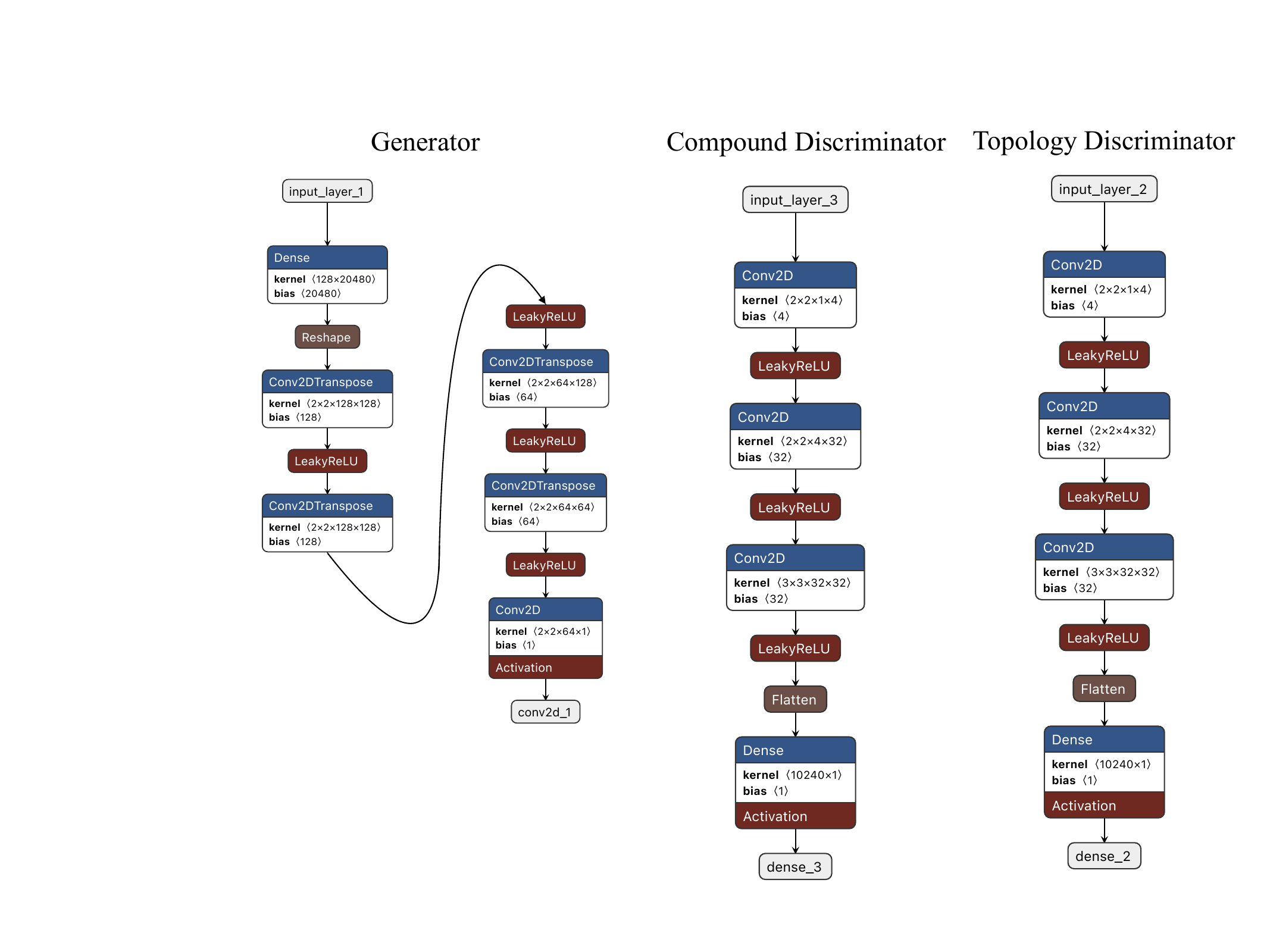}
    \caption{Architecture of generator network, crystal structure discriminator used in stage I of training and topology discriminator used in stage II of training.  }
    \label{fig:Networks}
\end{figure*}

The generator and discriminators developed in this work utilize the Keras software package. The architecture of the generator model as well as the baseline discriminator for distinguishing real and generated crystal structures and the discriminator used in the first fine-tuning stage for determining topological classification are given in Fig. \eqref{fig:Networks}. In the final stage of fine-tuning the discriminator takes the form of the crystal graph convolutional neural network (CGCNN) for which details are available in Ref. \cite{cgcnn} as well as the universal machine learned interatomic potential, CHGNet\cite{deng2023chgnet}. All necessary code demonstrating the integration of these external networks as discriminators is available on the corresponding GitHub link\cite{github}.

\section{Predicted topological materials}
Three generated materials with promising evidence of non-trivial topology have been presented in the main body in Fig. 5. In this section we display four additional generated insulators displaying evidence of non-trivial topology. The seven total generated materials displayed in Fig. 5 and in this section represent the generated insulators for which the strongest evidence for non-trivial topology have been found. It is these seven materials from which we arrive at the conclusion that the fine-tuned network has achieved a success rate of $3.5\%$ in generating target compounds. 
\begin{figure*}
    \centering
    \includegraphics[width=16cm]{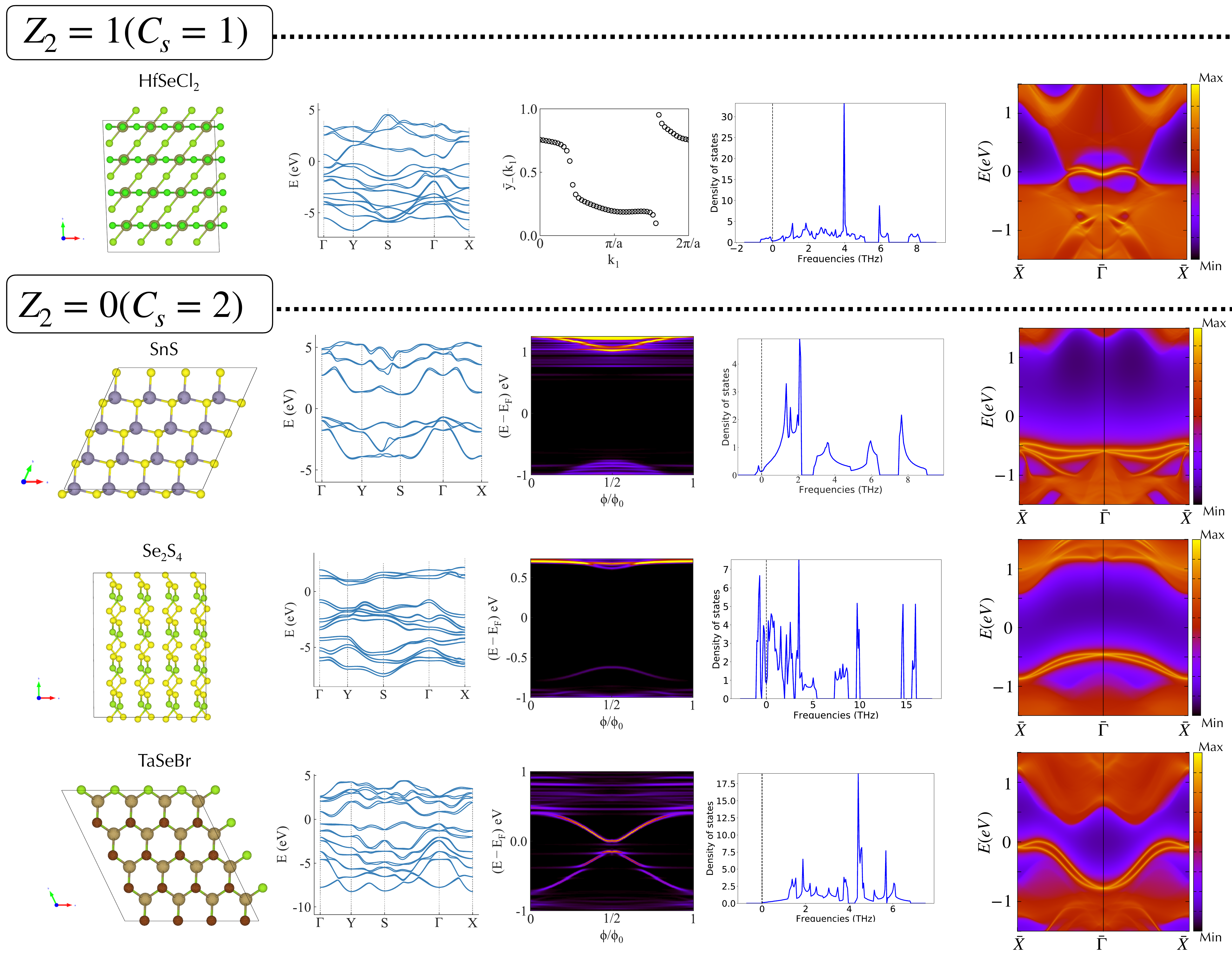}
    \caption{Additional topological materials generated by the fine-tuned network. Materials are separated by topological classification in terms of the spin-Chern number\cite{Prodan2009}. The crystal structure, band structure, evidence for topological classification and phonon density of states are shown in columns 1-4 respectively. Spin-resolved Wannier center charges are used as evidence for topological classification in the case of HfSeCl$_{2}$. For materials supporting $C_{s}=2$ we present the density of states on an inserted flux tube as a function of flux strength as evidence of non-trivial topology as this probe does not depend on the microscopic details of the preferred spin-quantization axis as detailed in the main body. }
    \label{fig:MatApp}
\end{figure*}
\par 
In Fig. \eqref{fig:MatApp}, we display the generated crystal structure, the band structure, evidence for non-trivial bulk topology (the spin-Chern number $C_{s}$) and the phonon density of states. For systems in which $C_{s}=1$ we have presented the spin-resolved Wannier center charges as evidence of the bulk topology. For systems in which $C_{s}=2$ we present the local density of states on an inserted flux tube as a function of the flux strength. This is done as, unlike in the case $C_{s}=1$, when $C_{s}=2$ it is possible to close and reopen the spin-gap, changing the preferred spin-quantization axis without closing the bulk gap. By contrast, examination of charge bound to an inserted flux tube is independent of the choice of spin-quantization axis. 
\par
\begin{figure*}
    \centering
    \includegraphics[width=16cm]{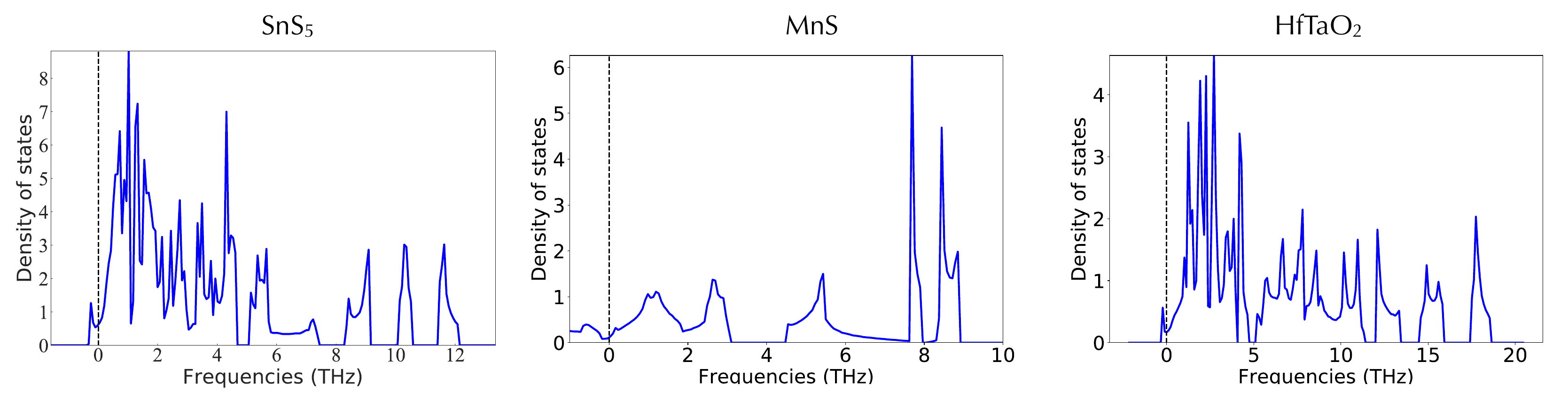}
    \caption{Phonon density of states for generated two-dimensional materials detailed in Fig. 5 of the main body. }
    \label{fig:MatPDos}
\end{figure*}

 
We further include the phonon density of states for the materials presented in the main body in Fig. \eqref{fig:MatPDos}. 

\bibliographystyle{apsrev4-1}
\nocite{apsrev41Control}
\bibliography{Ref.bib}